# Sublimation of orientated amino acid films for reliable, amplified piezoelectric performance

Ciaran O'Malley[1], Muhammad Usaid[1], Tara Ryan[1], Krishna Hari[1], Sarah Guerin[1,2]

[1]*Department of Chemical Sciences, Bernal Institute, University of Limerick, V94 T9PX, Ireland*

[2]*SSPC, The Science Foundation Ireland Research Centre for Pharmaceuticals, University of Limerick, V94 T9PX, Ireland*

**Abstract**

Biomolecular crystals, such as amino acids, peptides, and proteins, have emerged as potential next generation piezoelectric materials due to their low-cost, biocompatibility, eco-friendliness, and reduced permittivity versus ceramics. However, many challenges have limited their acceleration into mainstream sensing applications. Their natural self-assembly from saturated solutions into polycrystalline films reduces their effective piezoelectric output, and results in high variability within individual samples, and across sample batches when scaled-up. Here we validate the sublimation of a variety of amino acids onto conductive substrates as an effective technique for overcoming these challenges. This solvent-free crystallisation technique results in polycrystalline films with uniformly orientated crystals, and a resulting piezoelectric response that exceeds that of single crystals. We report a maximum piezoelectric response of 9.6 pC/N in films of L-Valine, which matches the predicted single crystal response, and a maximum voltage output of 4.6 V. For L-Methionine and L-Valine sublimated films the material properties are consistent at all points on the piezoelectric films and repeatable across any number of films grown under the same conditions.

**Introduction**

Biomolecular crystals are a general class of natural or synthetic materials including amino acids, peptides, proteins and viruses that display the ability to interact with biological systems, i.e. biocompatible[1]. This class of materials display functionality for a wide range of applications including pharmacology and drug delivery[2-4], crystallisation aids[5, 6] and piezoelectric sensors[1, 7]. While various methods for the production of biomolecular crystals are available, solution-based methods are by far the most dominant [8-11]. Methods for crystal growth and screening can be broadly characterised into solvent and non-solvent-based methods.

Solvent based methods can be an easy and reliable method for biomolecular crystallisation, they are not without their disadvantages. Persistent solvate formation and poor-quality crystal growth can hinder characterisation of crystalline systems while production and screening of crystalline systems can generate large quantities of solvent waste, especially on an industrial scale[12-14].

Crystallisation by evaporation remains by far the most dominant method to produce biomolecular crystals on a laboratory scale. Based on this strategy there are numerous examples of biomolecular crystals and films reported in the literature[5, 9, 15, 16]. While extremely useful for providing large single crystals required for diffraction studies, crystals produced in this method can suffer from persistent solvate formation[12, 13, 17] and crystallisation is limited by conformer solubility in the respective solution which can lead to poor crystal growth[18]. Crystal growth can also vary from several days to several months depending on the volatility of the solvent used. The production of bulk quantities using evaporative crystallisation can also produce a large amount of undesirable organic solvent waste whilst in terms of piezoelectric and optical response generation, crystalline orientation proves problematic to control[19].

Sublimation is defined as the direct transition of a solid to a vapour without passing through a liquid phase[20]. It is estimated that two-thirds of all organic compounds are sublimable[21] and while the use of sublimation has gained increasing attention recently for the growth of organic compounds, the use of sublimation for the preparation of biomolecular crystals and films is still in its infancy with a sparse number of reports in the last number of years[22-30]. Sublimation as a technique for crystal growth offers several advantages over traditional solvent based techniques, being a solvent free 'green' alternative, offering good polymorph[31-33] and morphology control[30, 34] by adjustment of the crystallisation driving force and industrial scale operation[35]. More often the use of a vacuum or an inert atmosphere has been employed to prevent oxidation of samples and to increase the sublimation rate. In such methods, close control of the sublimation rate and the temperature of the desublimation surface is critical in the quality of crystals grown[18]. Low nucleation rates increase crystal size by reducing the available area to condensate on. Low temperature gradients employed and clean glassware aids in producing high quality crystalline layers. A number of methods and experimental techniques have been described for the growth of crystals and powder films from the gas phase. Ranging from micro-spacing in air sublimation[22, 23], transpiration methods[24, 25], low temperature sublimation in vacuo (LTGSV)[18] and sealed tube under reduced pressure[36, 37].

What has been missing to date in the effort to accelerate the uptake of biomolecular crystals as high-performance commercial piezoelectrics, is a technique for growing polycrystalline assemblies in a repeatable manner with minimal cross-sample variation. Currently no technique orientates individual crystals within a film in a way that maximises their piezoelectric output versus single crystals. This is compounded by the fact that a large number of biomolecular crystals grow in monoclinic space groups that tend towards needle or plate morphologies that are difficult to align perpendicularly between the device electrodes.

In this work, we control the self-assembly of a number of amino acids by subliming them onto copper substrates. These films are characterised for their longitudinal piezoelectric response and made into simple three-layer disc-shaped actuators for output voltage measurements. The mechanical strength and directional self-assembly are presented via compression testing and various microscopy techniques. This characterisation and analysis is performed over a number of samples to demonstrate the reliability and repeatability of this technique for high-performance piezoelectric sensing and energy harvesting.

**Methods**

i.   Crystal Growth

The Sublimation Apparatus was constructed in house using commercially available materials as shown in Figure 1. Samples were sublimed in 150 x 24 mm test tubes connected to an open vacuum system with the tubes suspended within a heating bath controlled by an external heating system. Polycrystalline layers were deposited under reduced pressure on a 1.8cm diameter conductive substrate suspended 1cm above the starting material.

ii. Single-Crystal X-ray Diffraction

A Bruker D8 Quest fixed-χ single-crystal diffractometer equipped with a sealed-tube X-ray source that delivers Mo Kα (λ = 0.71073 Å), a TRIUMPH monochromator, a PHOTON 100 detector, and a nitrogen-flow Oxford Cryosystem attachment was used to collect X-ray diffraction data at 150K[38]. Unit cell determination, data reduction, and absorption correction (multiscan method) were conducted using the Bruker APEX3 suite[38]. The crystal structures were solved using ShelxT and refined using Shelxl within the Oscail package[39-41]. CIF files can be obtained free of charge atwww.ccdc.cam.ac.uk/conts/retrieving.html or from the Cambridge Crystallographic Data Centre, Cambridge, UK with the REF codes LMETOB12, LVALIN05 and LEUCIN02

iii. Powder X-ray Diffraction

X-ray powder patterns were recorded using a Panalytical Empyrean diffractometer with Cu Kα radiation (1.54060 Å, 40kV, mV). Diffraction patterns were collected in θ–2θ configuration at room temperature in the range 10–50° using a step size of 0.002°.

iv. SEM

SEM images were taken on a Hitachi SU-70. Crystals were coated in a 5 nm gold film prior to imaging, using a K550 Emitech Sputter Coater, with a coating current of 20 mA.

v. Optical Microscopy

Optical images were taken using an Olympus BX51 Light Microscope connected to an Olympus SC50 Digital Camera.

vi. Piezoelectric Measurements

Longitudinal piezoelectric constants were measured using a commercial PiezoTest $d_{33}$ meter, with an accuracy of 0.01 pC/N (pm/V)

vii. Electrical Testing

The electrical testing was done through a systematic procedure to ensure the reliability and scalability of the crystals. As the crystals was sublimated on the copper foil which acts as the bottom electrode, So for the top electrode the same copper tape was used and was princely cut into the circular piece of same diameter as the crystal. It was attached to the samples and prior to the wire attachments, the top and bottom electrodes surface was prepared for soldering to remove the dirt or any oxidation layers on it. Isopropyl alcohol was used for the cleaning purpose and a smooth layer of rosin flux was used to ensure smooth soldering and wettability.

For the electrical connections AWG 28 copper wires were attached to the crystals carefully using a lead-free solder. After strongly bonding the wires, the digital oscilloscope with high sampling rate was used to observe the voltage metrics of the crystals.

viii. Compression Testing

A Mecmesin Multi-test 2.5dv (Mecmesin, West Sussex, UK) device was used for compression testing to measure the macroscale strength of the samples. A sample presser of 2 mm was used, the sample height was set to 2 mm, and a force of 50 N was applied. The strain was allowed to reach 100 % and when this was plotted on a stress-strain curve it showed multiple sharp spikes, corresponding to cracks in the polycrystalline material before a final absolute fracture point is reached. The initial linear region until the first crack was taken and plotted. The slope of these lines tells us Young's modulus for each sample, or for anisotropic crystals, the elastic stiffness in the direction of the applied force.

**Results**

Films of L-Methionine, L-Valine, L-Leucine and DL-Alanine were grown via sublimation using an in-house apparatus as described in methods. 100 mg of sample was deposited in sublimation tube and heated between 160-200 °C for 12-96 hours. Films were then harvested from the tube. For optimal deposition, sublimation tubes were immersed to a depth of 1cm in heating oil with the deposition surface suspended just below the heating surface to direct deposition onto conductive substrate. The temperature observed at the sublimation surface and along the sides of tube was observed to hold steady over time within +/- 0.5 °C of the target temperature with the desublimation surface observed as ca.4 °C lower than the sublimation surface. With the desublimation surface suspended away from the walls of the glass tube a natural fine temperature gradient could be observed. Copper foil was used as a deposition surface as a conductive material with negligible thermal expansion over the heating ranges used. L-Methionine and L- Valine crystallised in the monoclinic $P2_1$ space group with the DL-Alanine in the orthorhombic $Pna2_1$ space group both of which are non-centrosymmetric and so elicit a nonzero piezoelectric response. Polymorphism was determined by PXRD as shown in (Figures SI XX-XX) compared with the reported structures as reported in the Cambridge Structural Database (Refcodes DLALNI07, LMETOB12 and LVALIN05)

*Sublimation of L-Methionine films*

L-Methionine was sublimed at varying temperatures between 160 and 200 °C for a period of 12 hours. A steady increase in deposited film mass from 5 to 68 mg with temperature could be observed correlating with an increased thickness and film density as seen in Figure S1. Samples were shown to be very replicable with little variation of weight gain between samples at a given temperature. Piezoelectric response showed a gradual decline with increased film density from a high of 6.57 pC/N at 6mg to 3.31 pC/N at 68 mg. The maximum response matches well with the predicted $d_{22}$ value of L-Methionine single crystals (6.9 pC/N)[42]. These responses showed a low standard deviation both between and across samples as evidenced in Table S1 SEM showed an increasing loss of control over film morphology with increasing temperature impacting the piezoelectric response.

Samples were sublimed between 0.5 and 12 hours at 200 °C. A general increase in mass gain from 4 to 68 mg with increasing time could be observed correlating with an increased thickness in film density as seen in Figure S2. The deviation of weight gain between samples was close for up to 2 hours with increasingly deviating after with increased time scales correlating to erratic secondary sublimation process at the condensation surface. This can also be seen in the piezoelectric responses where no significant correlation could be made due to high deviations between samples. This can be evidenced in microscopy images (Figure S4) where samples show varying degrees of thickness and consistency

Samples were sublimed between 12 and 96 hours at 160 °C. A general increase in mass gain from 5 to 18 mg with increasing time could be observed correlating with an increased thickness in film density as seen in Figure S8. Like samples at 200 °C, Increased time scales showed increasing deviation of mass gain with samples consistent up unto 24hrs. piezoelectric responses were generally consistent across samples and showed a gradual decrease in max response from 6.57 pC/N after 12 hours to 5.44 pC/N after 72 hrs

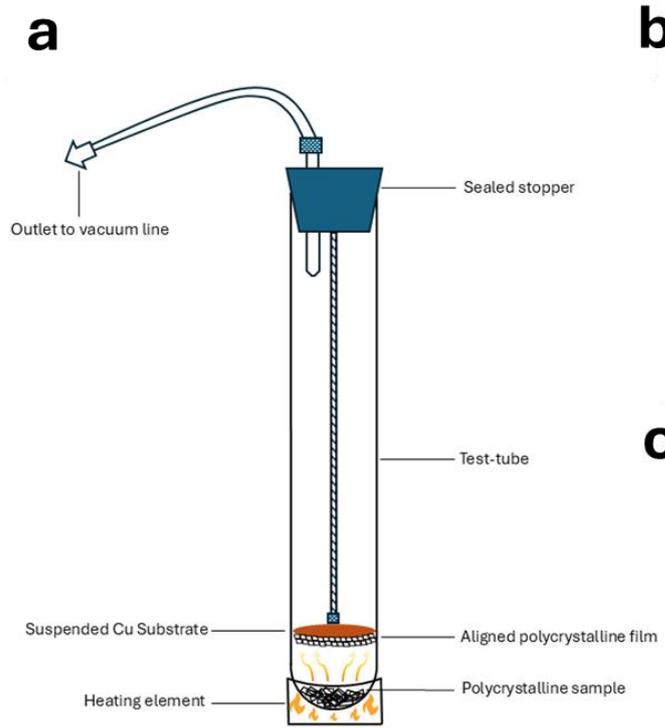
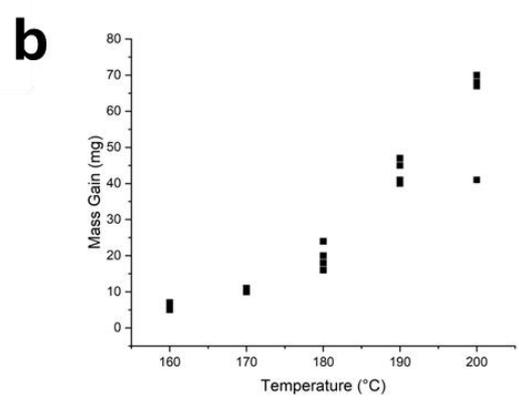
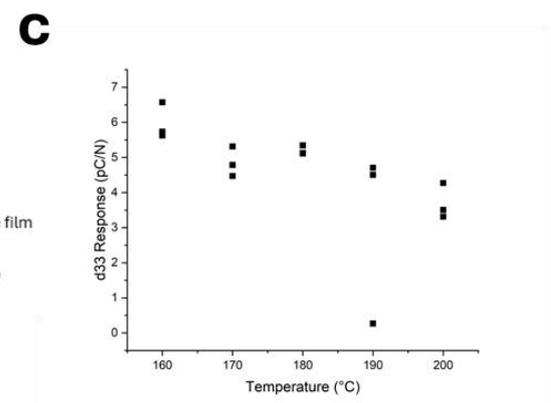
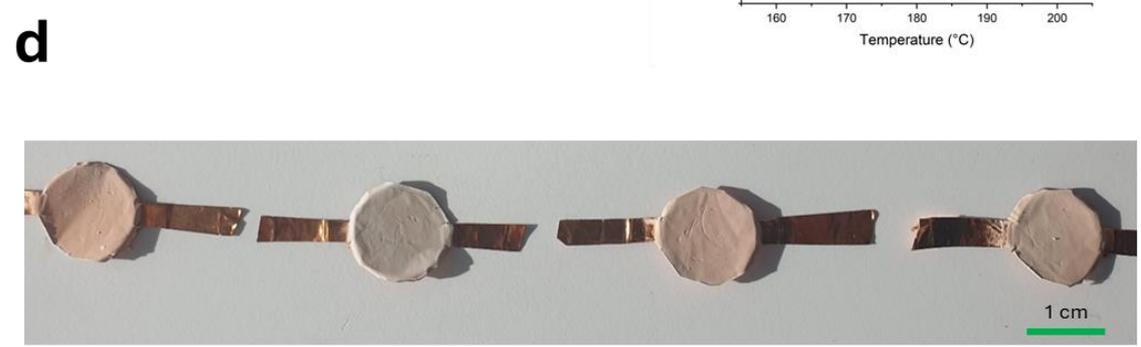
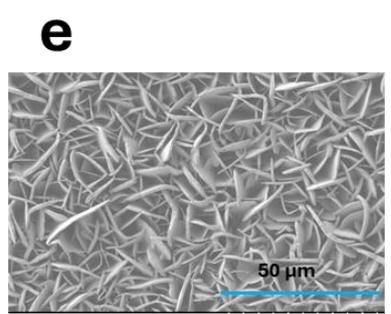
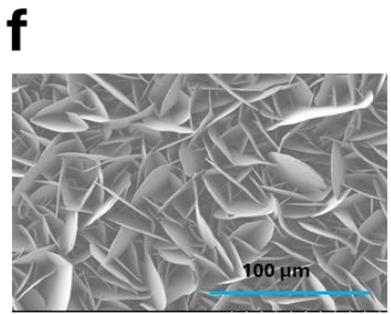
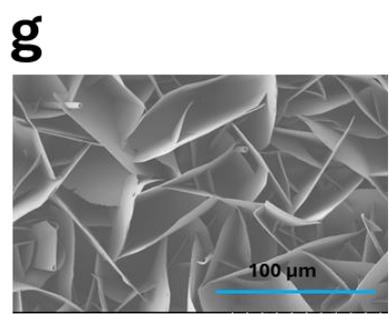
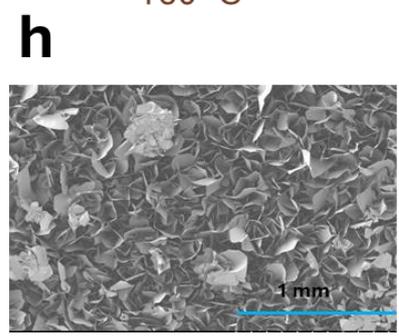
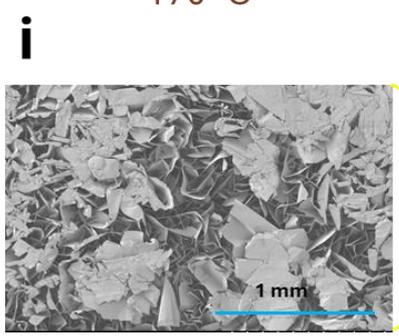
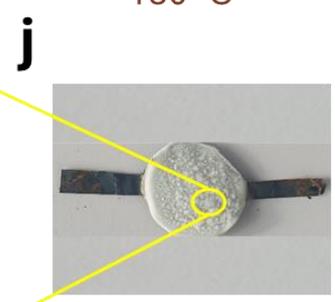

**Figure 1 Sublimation of orientated L-Methionine crystal films a)** Simplified schematic of sublimation system **b)** Plot of mass gain versus temperature keeping time constant for L-methionine **c)** Plot of longitudinal piezoelectric response ($d_{33}$) versus temperature while keeping sublimation time constant for L-methionine **d)** Image of four L-Methionine films sublimated onto a circular copper substrate in identical conditions at 160 °C that exhibit single crystal piezoelectric responses. SEM image of L-Methionine crystal film surface when sublimated at a temperature of **e)** 160 °C **f)** 170 °C **g)** 180 °C **h)** 190 °C **i)** 200 °C **j)** Image of thick L-Methionine film grown at 200 °C, corresponding to the magnified surface in panel **i**.

*Sublimation of L-Valine films*

L-Valine was sublimed at varying temperatures between 160 and 200 °C for a period of 12 hours. An erratic mass gain was observed across temperatures with no correlation observed. The mass deviation between samples was large ranging from 34 to 59 mg at 180 °C with a mass range of 31 mg to 58 mg at temperatures either side of this. Samples in general could be observed to have a wide deviation in middle temperature ranges at 180 °C with samples either side of this range exhibiting a lower deviation in the form of a bell curve (Figure 3a). This is observed to be due to the initial film morphology being tightly controlled at low temperatures with increasing loss of control with secondary sublimation at higher temperatures (Figure 3b). At 160 °C it can be seen a much lower mass gain was observed (between 10 and 20 mg) with morphology effects preserved. After this point an equilibrium is reached between the rates of sublimation and desublimation resulting much the same observed mass gain between 170 and 200 °C

When samples were sublimed at a constant temperature of 200 °C, and the time of sublimation was varied between 0.5 and 12 hours at mass gain resembling a spring parachute model was observed. A maximum mass gain was observed between 2 and 4 hours (91 mg) with a gradual decline to 34 mg observed with further increases in sublimation time. Piezoelectric responses showed an initial erratic behaviour with consistently high results of 7.74-8.15 pC/N achieved after 4 hours. This can be attributed to a *quasi*-annealing effect where initial rapid crystallisation results in uncontrolled morphology with subsequent desublimation orientating the deposited film.

Finally, L-Valine samples were sublimed between 12 and 96 hours at a lower temperature of 160 °C. A general increase in mass gain from increasing time could be observed correlating with an increased thickness in film density. While mass gains at longer timescales were on average higher (up to 88 mg), the standard deviation between samples significantly increases after 24 hours from +/- 8 mg at 24 hrs to +/- 32 mg after 36 hrs. In contrast, the piezoelectric responses, while lowering slightly over longer timescales, become more consistent as orientation effects dominate (+/-2.2 pC/N after 96 hrs).

*Sublimation of L-Leucine films*

L-Leucine films showed higher cross-sample and inter-variability than other amino acids in this study. over a constant time of 12 hours but produce a maximum repeatable response of 9.1 +/- 0.09 pC/N at 150 degrees. Prior to this the highest longitudinal response measured in L-Leucine films grown from solution was only 1.9 pC/N, with the highest sublimated values in this work being near-identical to the single crystal DFT-predicted response of 8.6 pC/N. At 130 degrees the sample with the lowest mass gain of 46 mg had a low response of 0.9 pC/N with a high inter-sample standard deviation of 0.37 pC/N. The samples with higher mass gains between 53 and 57 mg had repeatable responses between 3.4 and 4 pC/N. The two samples produced at 160 degrees had responses of 7.16 and 5.43 pC/N, corresponding to their mass gains of 40 and 28 mg respectively. However, at 170 degrees the mass gain is inversely proportional to the average longitudinal piezoelectricity, with responses of 5.48 and 3.37 pC/N corresponding to mass gains of 72 and 81 mg respectively.

When keeping the temperature constant at 200 degrees and varying the time, the highest measured response was recorded for a time of 4 hours and a mass gain of 69 mg. For the same conditions the response decreased with increasing mass gain to a minimum of 2.9 pC/N in a sample of mass 79 mg. The same experiment conducted at 160 degrees resulted in films with consistently high responses distributed between 3.4 and 8.72 pC/N, with the most consistent responses at 8 hours.

*Crystal Orientation*

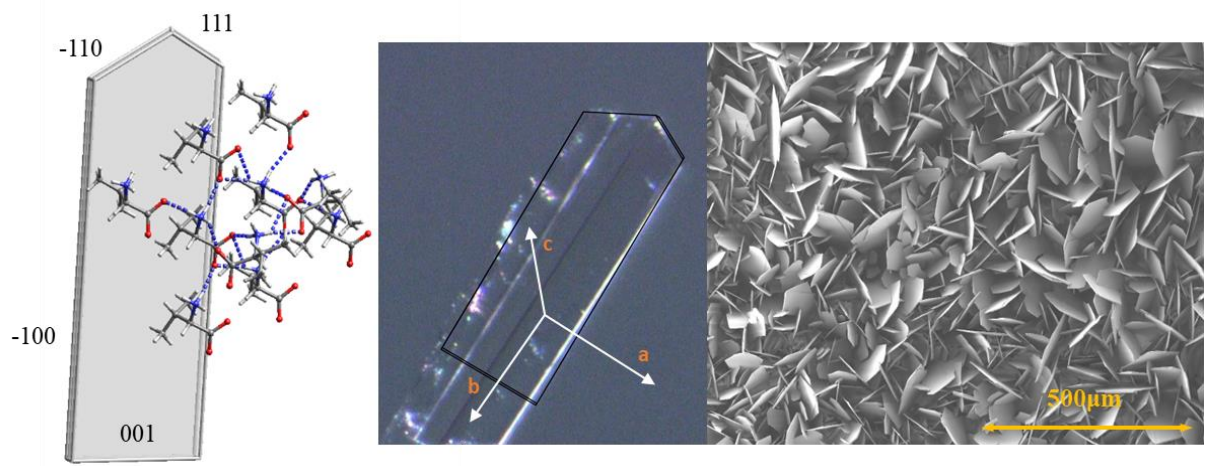

**Figure 2 Morphology of orientated L-Valine Crystal films a**) Experimental face indexed morphology of L-Valine crystals overlayed on growth pattern   **b)** Indexed crystal of L-Valine from sublimation showing growth direction in b-axis **c)** SEM of bulk L-Valine growth showing orientated needles in the sublimated crystal film

Sublimated crystalline films of L-Valine, L-Methionine and L-Leucine showed a high degree of crystalline orientation in the SEM images. By harvesting crystals produced via sublimation we were able to face index these products to determine the preferred orientation of these films. All three systems crystallise in monoclinic space groups, and it can be seen that the crystals are naturally orientated along the b-axis in each case corresponding to the main hydrogen bond driven growth direction. This corresponds to the observed experimental $d_{33}$ values of the sublimated films matching well with the calculated values for $d_{22}$ of each crystal system with the $d_{22}$ tensor running along the b-axis.

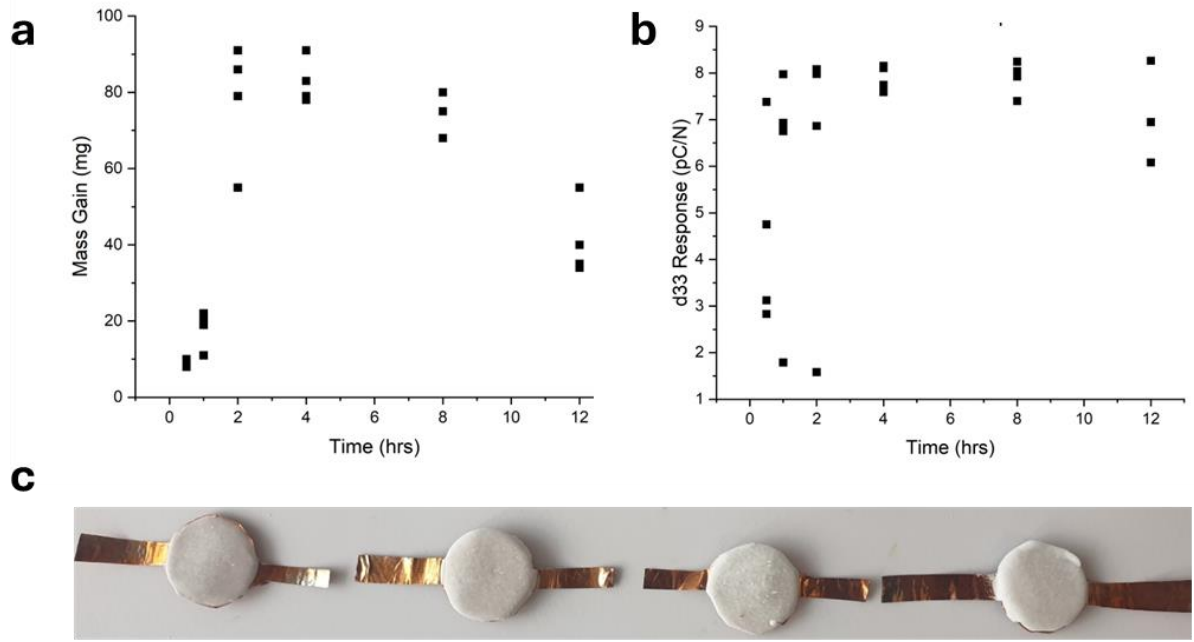

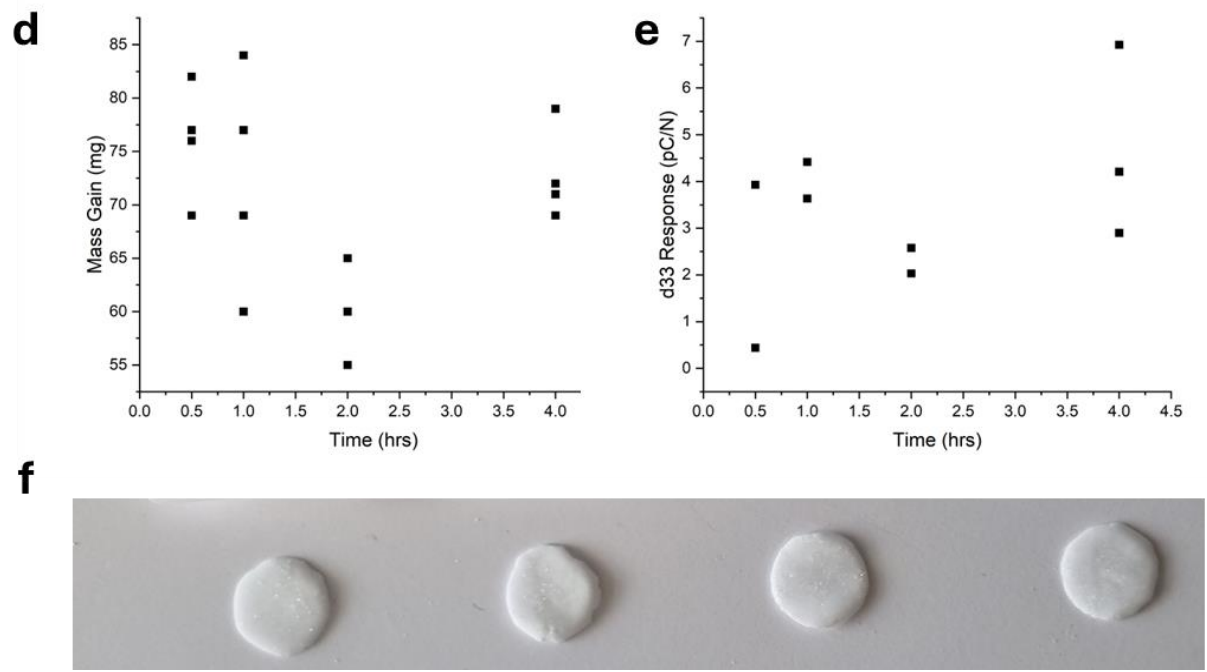

**Figure 3 Sublimation of orientated L-Valine and L-Leucine crystal films a)** Plot of mass gain versus sublimation time while keeping temperature constant for L-valine **b)** Plot of longitudinal piezoelectric response ($d_{33}$) versus sublimation time while keeping temperature constant for L-valine **c)** Image of four L-valine films grown in identical conditions on a circular copper substrate that show high piezoelectric response **d)** Plot of mass gain versus sublimation time while keeping temperature constant for L-valine **e)** Plot of longitudinal piezoelectric response ($d_{33}$) versus sublimation time while keeping temperature constant for L **f)** Image of four L-leucine discs grown in identical conditions on a circular copper substrate that show high piezoelectric response. Films detached from copper substrate after growth to become stand-alone piezoelectric elements.

*Energy Harvesting Tests*

To evaluate the energy harvesting capabilities of the films and the repeatability of the voltage outputs under an applied force, a second piece of copper tape was applied as a top electrode of the same diameter. The simple sandwich-type piezoelectric sensor, similar to our previous work, was connected to a multimeter via soldered wires and subjected to tapping forces. Across all of the film types (both chemistry and sublimation conditions), the highest measured response was 8.4 V peak-peak This was measured in a sample of L-Methionine grown at 200 for 4 hours. For a set of multiple samples of L-Methionine grown simultaneously under identical conditions both the longitudinal piezoelectricity and output voltage could be replicated as 5.69 ± 0.23 pC/N and 1.46 ± 0.13 V respectively. Figure 3a and b show the longitudinal piezoelectric response of the films as measured without a top electrode, and the corresponding voltage output with a top electrode. The full data is shown in Supplementary Tables S11-S13.

*Mechanical Testing*

To determine the effective elastic stiffness of the films as well as their mechanical response to an applied force, compression testing was carried out on samples for each set of conditions. Figure4c and d show the resulting stress-strain curves of L-Methionine and L-Valine respectively. The measurements clearly show that varying the temperature and time of the sublimation process can produce films with a target flexibility. The elastic stiffness of L-Methionine Films varied from 1.3 GPa for films produced at Y, to 14.1 GPa for films grown under Z. This was observed visually in all three amino acids, with thinner films being highly bendable (Table S14 & S15). Figure 3e shows the stress-strain response up to 100% for representative films of L-Valine and L-Methionine, versus two of our recent works on polycrystalline discs formed via drop casting into silicone molds. For gamma glycine and the L-Lysinium.S.Mandalate.5H$_2$O cocrystal (**1**), the thick and more brittle nature of the as-formed crystals results in a jagged stress-strain response, with multiple cracks forming before the polycrystalline assembly fully breaks. The sublimated films contrastingly, demonstrate a near-continuous linear response up to 100% strain, making them more suitable for high-force sensing applications than our previous polycrystalline discs.

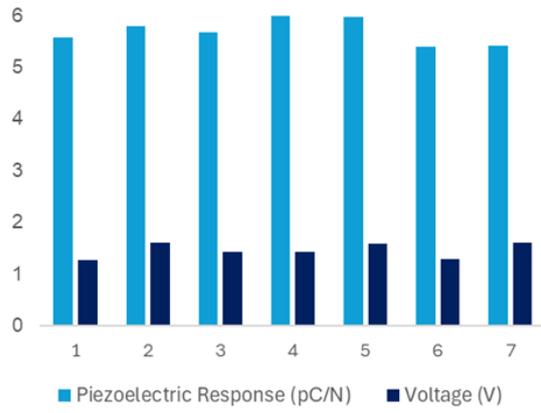
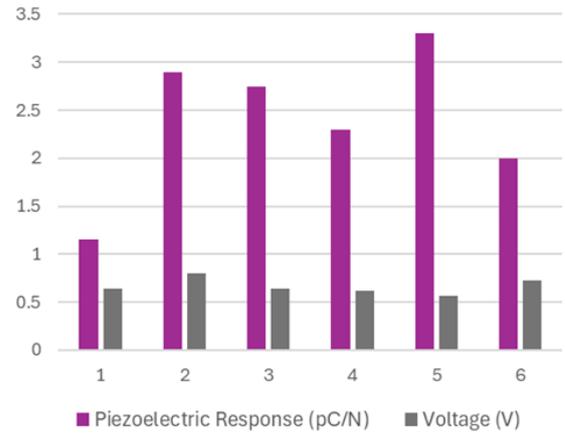
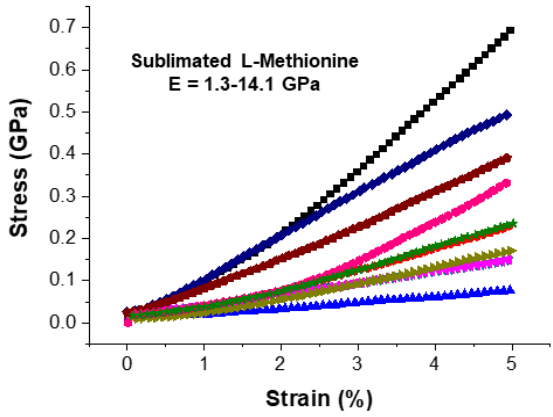
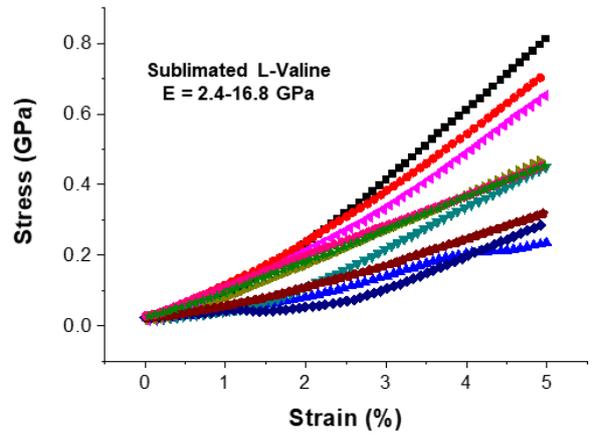
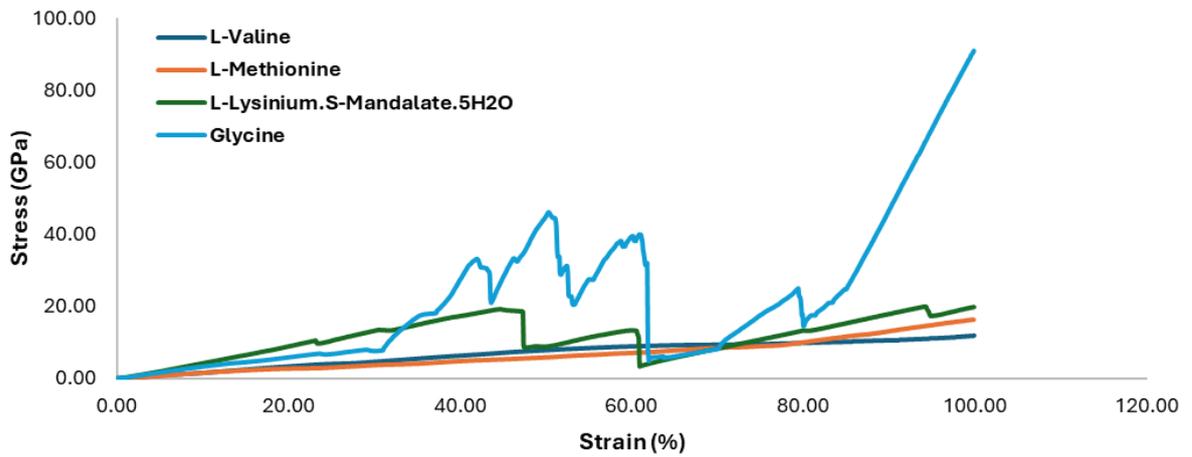

**Figure 4 Electromechanical properties of Sublimated Amino Acid Films a**) Highly consistent longitudinal piezoelectric response and voltage measurements for a series of samples of L-Methionine prepared at 200°C for 4 hours **b**) Fluctuations in longitudinal piezoelectric response and voltage output in films of L-Valine **c**) Linear regions of the stress-strain curves for L-methionine sublimated across a variety of conditions with the extracted range of elastic stiffness values **d**) Linear regions of the stress-strain curves for L-methionine sublimated across a variety of conditions with the extracted range of elastic stiffness values **e**) Comparison of the 'full' stress-strain curves (up to 100% strain) for two sublimated films versus two of our recent molecular polycrystalline films fabricated using a macroscopic molding technique

**Conclusions**

In this work, we have sublimated ordered, polycrystalline films of three piezoelectric amino acids: L-Methionine, L-Valine, and L-Leucine. The crystals self-assembled with their polar axis consistently aligned perpendicular to the growth substrate, resulting in a maximised piezoelectric response, and a significantly reduced cross-sample standard deviation versus conventional crystallisation techniques. The response across the sample also had low variability, a stark contrast to previously drop-cast films. This is the first time that polycrystalline amino acid films have demonstrated repeatable performance that matches the response of a single crystal. The films demonstrate significantly improved mechanical performance versus films grown from solution, with a large elastic region that makes these materials suitable for high force energy harvesting applications. This work highlights that sublimation is the ideal technique for scaling up production of sustainable molecular piezoelectrics, particularly for prominent applications in medical devices and energy harvesting.

**Acknowledgements**

C.O.M., M.U., T.R., and S.G. are funded by the European Union under ERC Starting Grant no. 101039636. KH and S. G. would like to acknowledge funding from Research Ireland under grant number 21/PATH-S/9737 and S.G acknowledges Research Ireland award number 12/RC/2275_P2.

# Supplementary Information

Sublimation of orientated amino acid films for reliable, amplified piezoelectric performance

Ciaran O'Malley[1], Muhammad Usaid[1], Krishna Hari[1], Tara Ryan[1], Sarah Guerin[1,2]

[1]*Department of Chemical Sciences, Bernal Institute, University of Limerick, V94 T9PX, Ireland*

[2]*SSPC, The Science Foundation Ireland Research Centre for Pharmaceuticals, University of Limerick, V94 T9PX, Ireland*

|  | **Description** | **Page** |
|---|---|---|
| Table S1 | Sublimation crystallisation experiments at variable temperature for 12 hours for 100mg L-Methionine | 2 |
| Figure S1 | Sublimation crystallisation samples (diameter 18mm) prepared at variable temperature for 12 hours for 100mg L-Methionine. A 160°C, B 170°C, C 180°C, D 190°C and E 200°C | 3 |
|  |  |  |
|  |  |  |

**Table S1.** Sublimation crystallisation experiments at variable temperature for 12 hours for 100mg L-Methionine

| Temperature (°C) | Sample | Weight Gain (mg) | $d_{33}$ Response (pC/N) |
|---|---|---|---|
| 160 | 1 | 7 | 5.62 ± 0.44 |
|  | 2 | 6 | 6.57 ± 0.66 |
|  | 3 | 5 | 5.73 ± 0.48 |
| 170 | 1 | 11 | 5.32 ± 0.39 |
|  | 2 | 10 | 4.79 ± 0.14 |
|  | 3 | 11 | 4.47 ± 0.22 |
| 180 | 1 | 16 | 5.13 ± 0.10 |
|  | 2 | 24 | 5.11 ± 0.48 |
|  | 3 | 20 | 5.35 ± 0.12 |
| 190 | 1 | 47 | 4.51 ± 0.31 |
|  | 2 | 41 | 4.71 ± 0.37 |
|  | 3 | 45 | 4.27 ± 0.22 |
| 200 | 1 | 41 | 3.50 ± 1.15 |
|  | 2 | 68 | 3.31 ± 0.61 |
|  | 3 | 67 | 4.27 ± 0.46 |

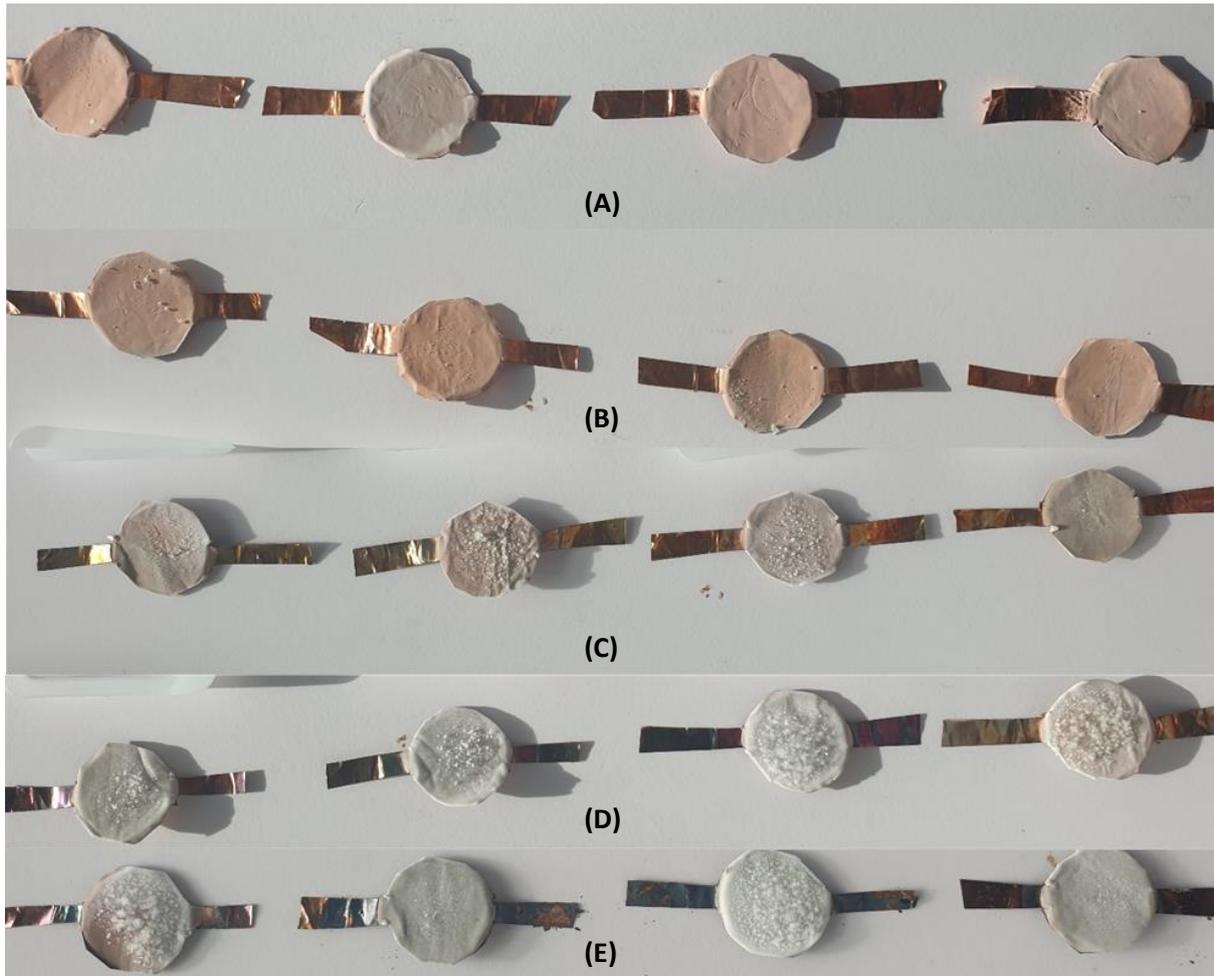

**Figure S1.** Sublimation crystallisation samples (diameter 18mm) prepared at variable temperature for 12 hours for 100mg L-Methionine. A 160°C, B 170°C, C 180°C, D 190°C and E 200°C

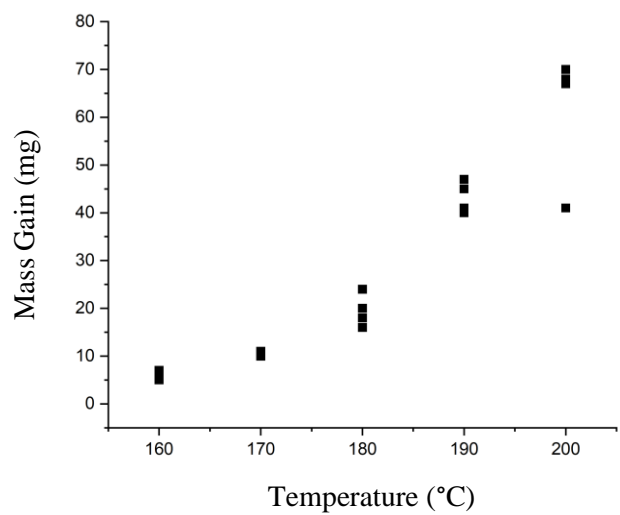

**Figure S2.** Plot of Temperature vs Mass gain for Sublimation samples prepared at variable temperature for 12 hours for 100mg L-Methionine.

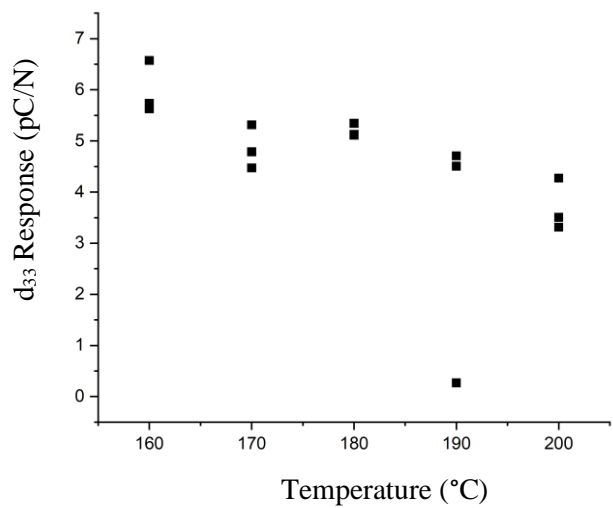

**Figure S3.** Plot of Temperature vs Piezoelectric response for Sublimation samples prepared at variable temperature for 12 hours for 100mg L-Methionine.

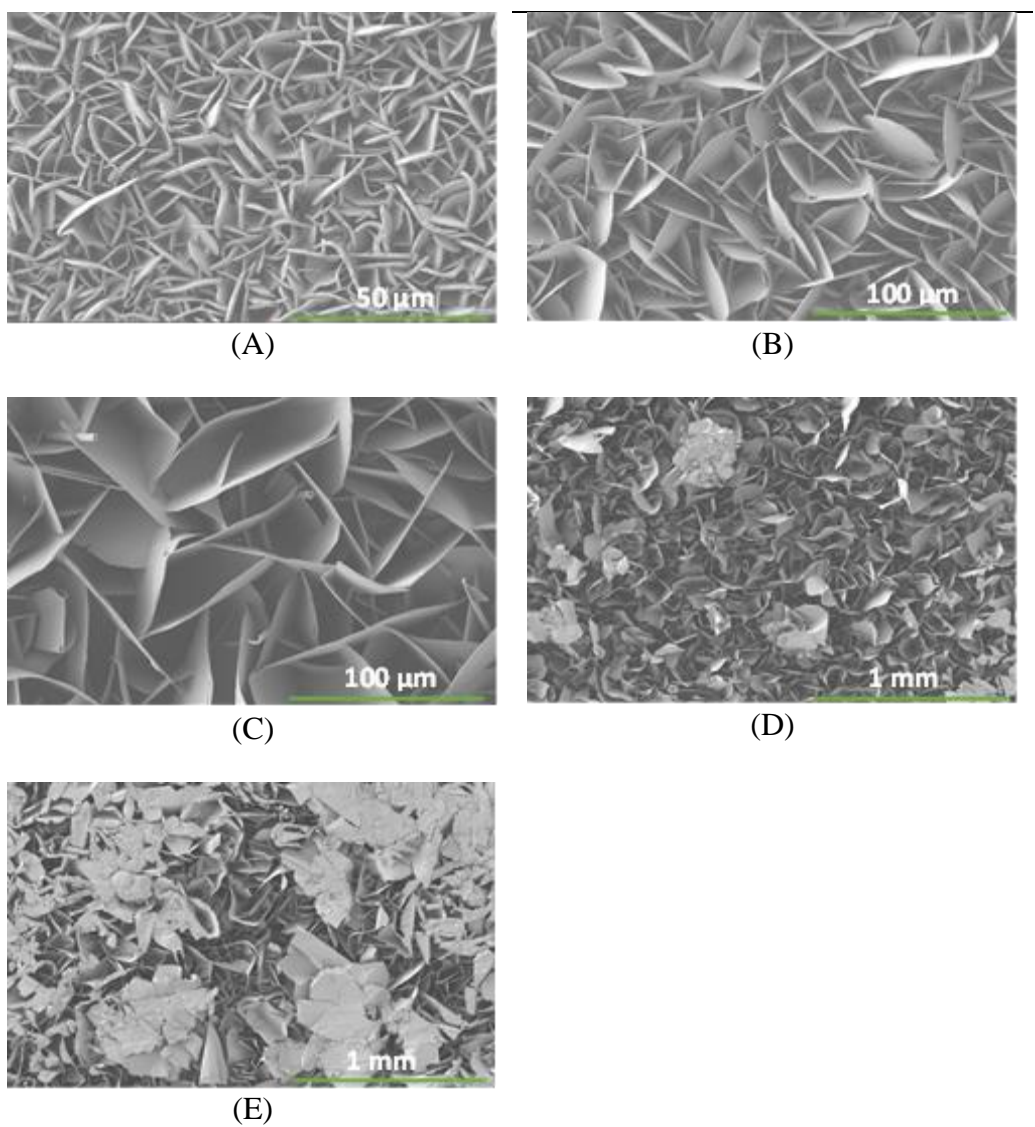

**Figure S4.** Sublimation crystallisation samples prepared at variable temperature for 12 hours for 100mg L-Methionine. A 160°C, B 170°C, C 180°C, D 190°C and E 200°C.

**Table S2.** Sublimation crystallisation experiments at variable time for at 200°C for 100mg L-Methionine

| Time (hrs) | Sample | Weight Gain (mg) | $d_{33}$ Response (pC/N) |
|---|---|---|---|
| 0.5 | 1 | 6 | 4.11 ± 0.36 |
| | 2 | 4 | 1.41 ± 0.15 |
| | 3 | 7 | 0.56 ± 0.15 |
| 1 | 1 | 9 | 2.77 ± 0.35 |
| | 2 | 12 | 4.56 ± 0.51 |
| | 3 | 9 | 2.9 ± 0.32 |
| | 4 | 10 | 4.25 ± 0.29 |
| 2 | 1 | 32 | 1.73 ± 0.44 |
| | 2 | 33 | 4.13 ± 0.26 |
| | 3 | 22 | 3.61 ± 0.41 |
| 4 | 1 | 48 | 5.78 ± 0.35 |
| | 2 | 47 | 0.89 ± 0.31 |
| | 3 | 30 | 2.56 ± 0.09 |
| 8 | 1 | 66 | 0.62 ± 0.27 |
| | 2 | 27 | 3.47 ± 0.37 |
| | 3 | 36 | 4.76 ± 0.23 |
| 12 | 1 | 41 | 3.51 ± 1.15 |
| | 2 | 68 | 3.31 ± 0.61 |
| | 3 | 67 | 4.27 ± 0.46 |

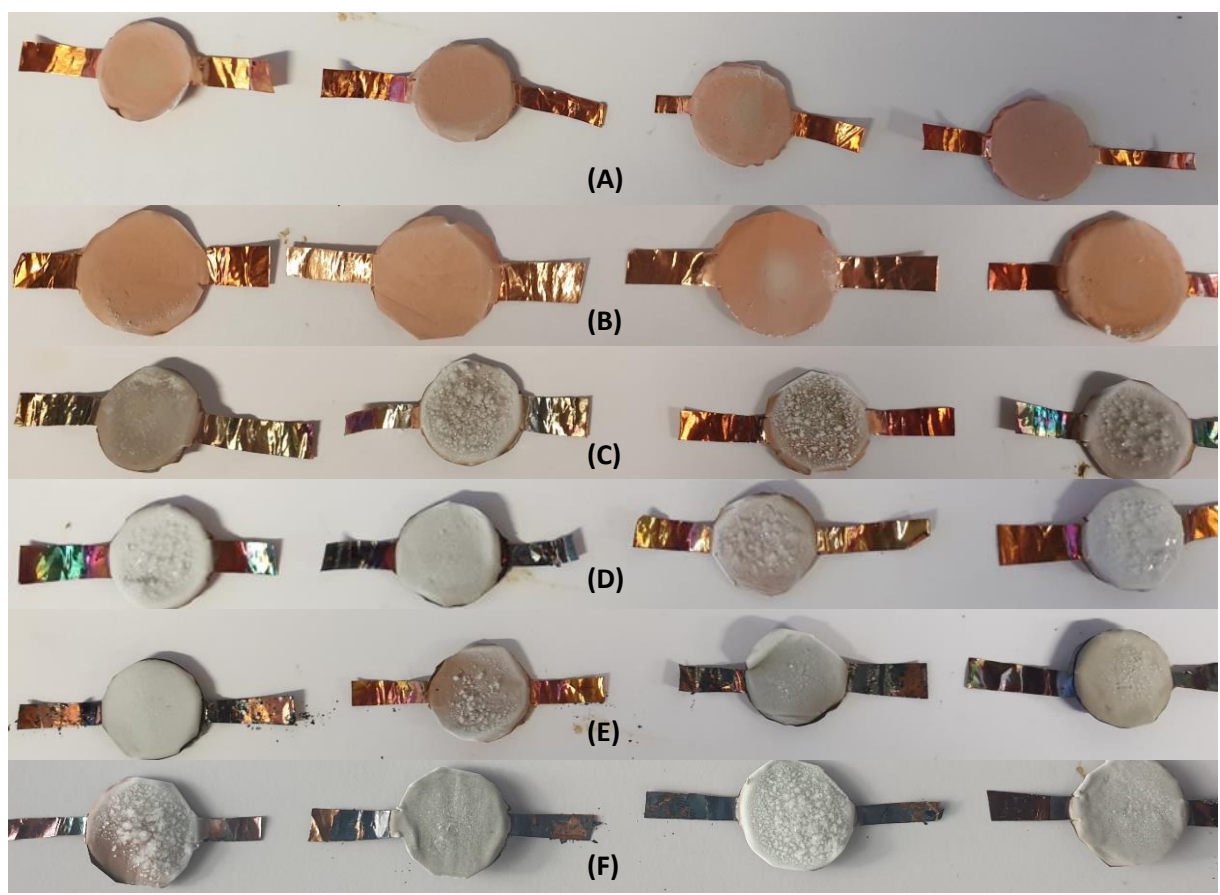

**Figure S5.** Sublimation crystallisation samples (diameter 18mm) prepared at variable time at 200° for 100mg L-Methionine. A 30mins, B 1hr, C 2 hrs, D, 4hrs, E 8hrs, F 12hrs.

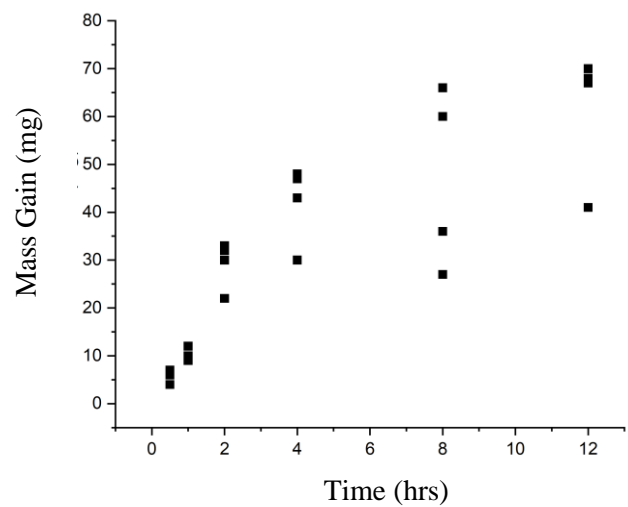

**Figure S6.** Plot of Time vs Mass gain for Sublimation samples prepared at variable time at 200°C for 100mg L-Methionine.

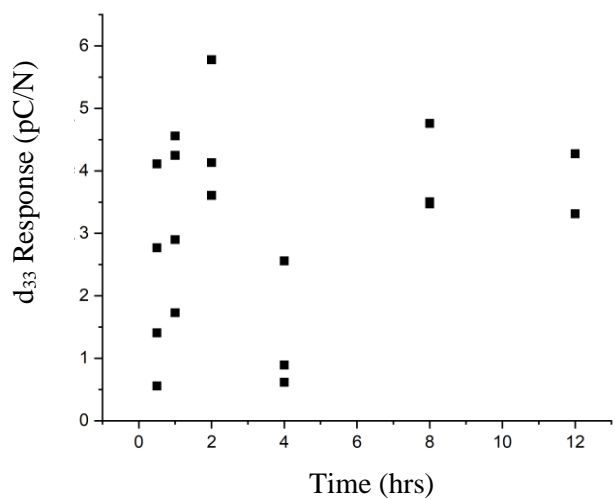

**Figure S7.** Plot of Time vs Piezoelectric response for Sublimation samples prepared at variable time at 200°C for 100mg L-Methionine.

**Table S3.** Sublimation crystallisation experiments at variable time for at 160°C for 100mg L-Methionine

| Time (hrs) | Sample | Weight Gain (mg) | $d_{33}$ Response (pC/N) |
|---|---|---|---|
| 12 | 1 | 7 | 5.63 ± 0.44 |
|  | 2 | 6 | 6.57 ± 0.66 |
|  | 3 | 5 | 5.73 ± 0.48 |
| 24 | 1 | 7 | 4.38 ± 0.15 |
|  | 2 | 6 | 5.20 ± 0.14 |
|  | 3 | 7 | 5.05 ± 0.29 |
|  | 4 | 5 | 4.55 ± 0.30 |
| 36 | 1 | 5 | 2.03 ± 0.42 |
|  | 2 | 7 | 3.21 ± 0.53 |
|  | 3 | 7 | 5.82 ± 0.22 |
|  | 4 | 14 | 4.99 ± 0.18 |
| 48 | 1 | 8 | 4.91 ± 0.20 |
|  | 2 | 5 | 3.82 ± 0.31 |
|  | 3 | 13 | 4.77 ± 0.27 |
|  | 4 | 14 | 5.34 ± 0.24 |
| 72 | 1 | 5 | 3.99 ± 0.42 |
|  | 2 | 12 | 4.69 ±0.41 |
|  | 3 | 16 | 5.44 ± 0.43 |
|  | 4 | 15 | 4.23 ± 0.29 |

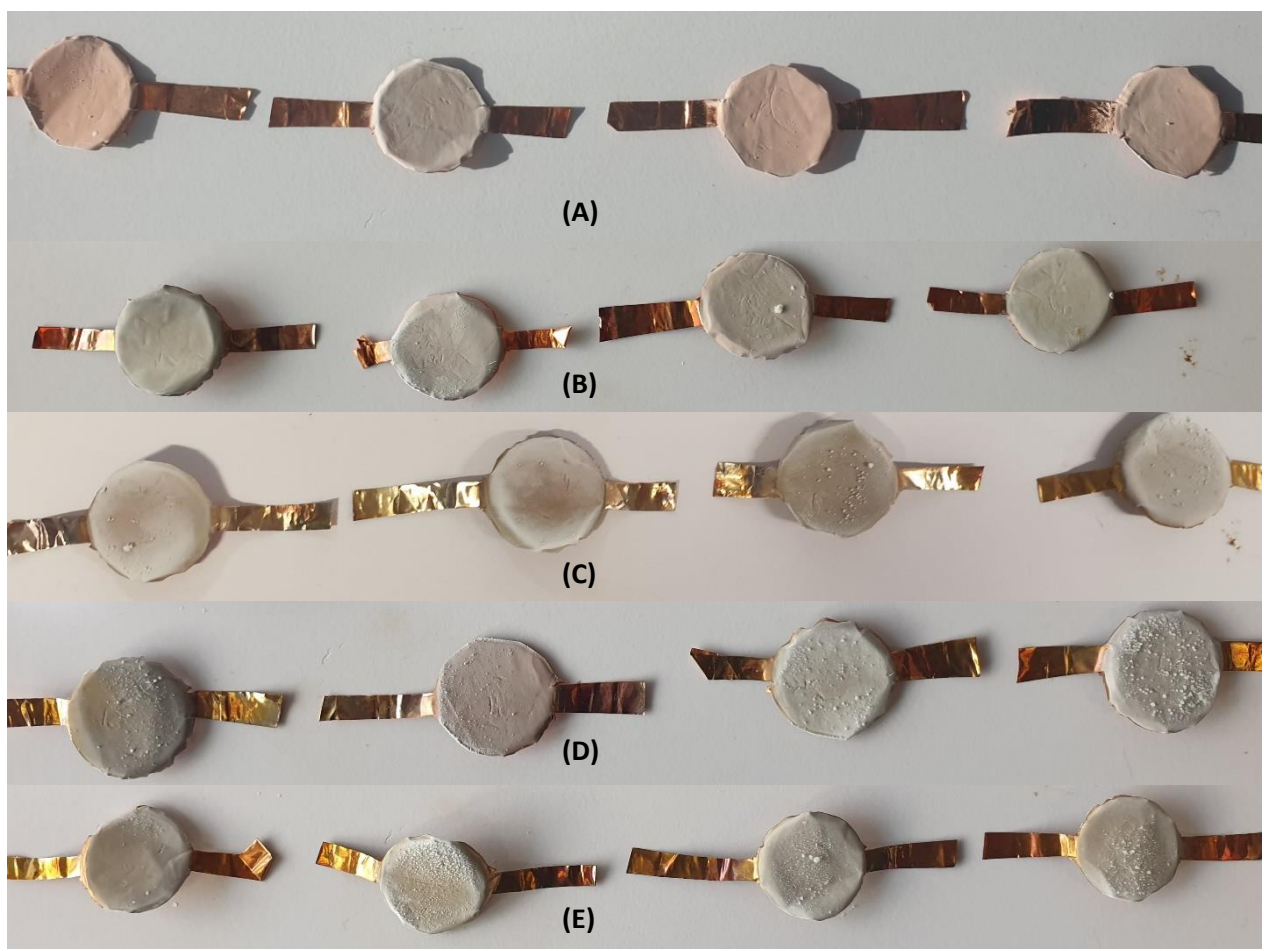

**Figure S8.** Sublimation crystallisation samples (diameter 18mm) prepared at variable time at 160° for 100mg L-Methionine. A 12hrs, B 24hrs, C 36 hrs, D, 48hrs, E 72hrs

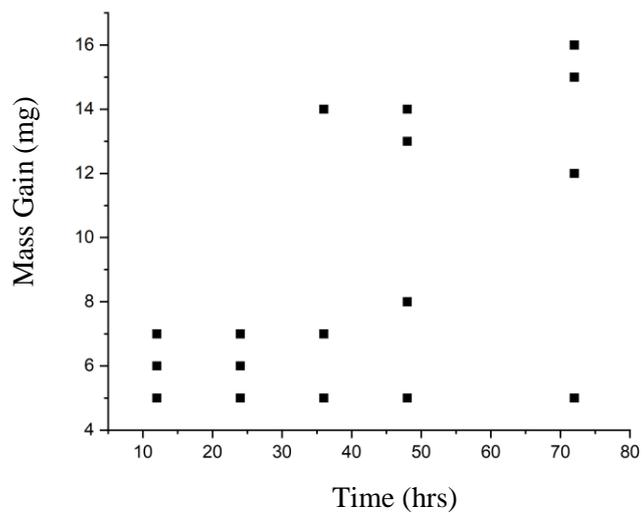

**Figure S9.** Plot of Time vs Mass gain for Sublimation samples prepared at variable time at 160°C for 100mg L-Methionine.

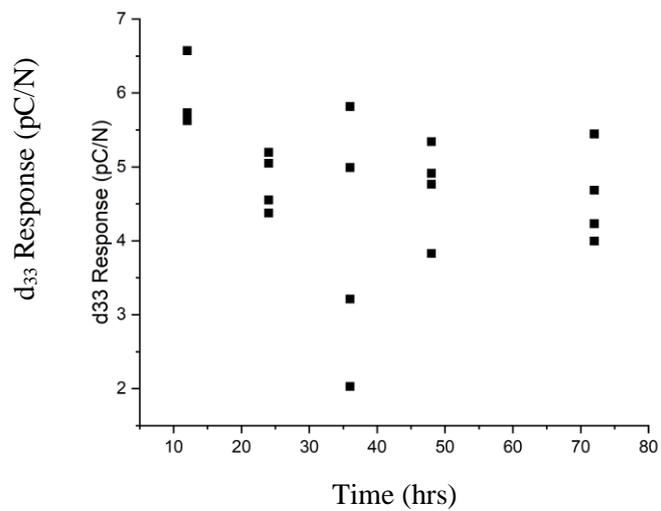

**Figure S10.** Plot of Time vs piezoelectric response for Sublimation samples prepared at variable time at 160°C for 100mg L-Methionine.

**Table S4.** Sublimation crystallisation experiments at variable temperature for 12 hours for 100mg L-Valine

| Temperature (°C) | Sample | Weight Gain (mg) | $d_{33}$ Response (pC/N) |
|---|---|---|---|
| 160 | 1 | 10 | 4.45 ± 0.75 |
|  | 2 | 17 | 5.05 ± 0.51 |
|  | 3 | 14 | 6.83 ± 0.35 |
|  | 4 | 20 | 5.87 ± 0.74 |
| 170 | 1 | 51 | 8.14 ± 0.23 |
|  | 2 | 36 | 7.63 ± 0.67 |
|  | 3 | 58 | 9.61 ± 0.27 |
|  | 4 | 39 | 8.50 ± 0.37 |
| 180 | 1 | 37 | 8.52 ± 0.61 |
|  | 2 | 59 | 8.62 ± 0.08 |
|  | 3 | 34 | 7.79 ± 0.48 |
| 190 | 1 | 43 | 5.41 ± 2.06 |
|  | 2 | 31 | 7.08 ± 0.17 |
|  | 3 | 44 | 6.79 ± 0.97 |
|  | 4 | 42 | 8.45 ± 0.50 |
| 200 | 1 | 55 | 6.95 ± 0.22 |
|  | 2 | 40 | 6.08 ± 1.57 |
|  | 3 | 34 | 8.27 ± 0.57 |

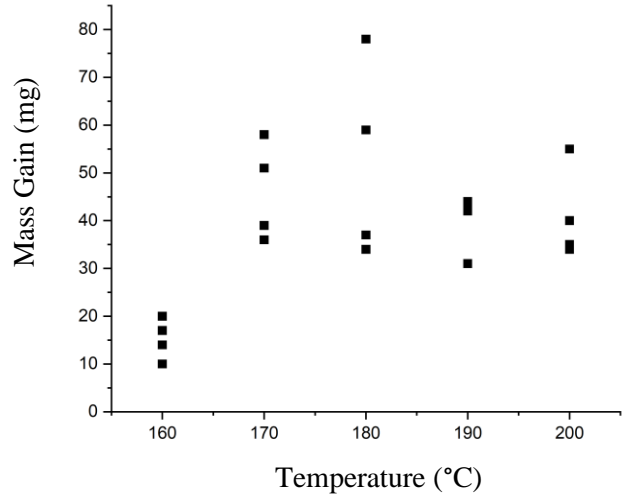

**Figure S11.** Plot of Temperature vs Mass gain for Sublimation samples prepared at variable temperature for 12 hours for 100mg L-Valine.

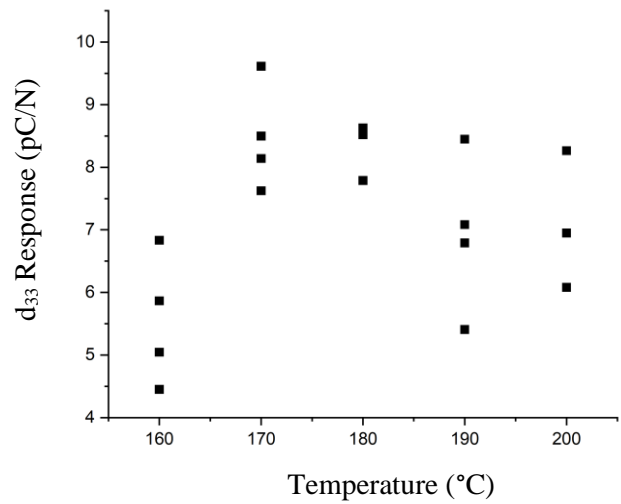

**Figure S12.** Plot of Temperature vs piezoelectric response for Sublimation samples prepared at variable temperature for 12 hours for 100mg L-Valine.

**Table S5.** Sublimation crystallisation experiments at variable time for at 160°C for 100mg L-Valine

| Time (hrs) | Sample | Weight Gain (mg) | $d_{33}$ Response (pC/N) |
|---|---|---|---|
| 0.5 | 1 | 10 | 2.83 ± 1.49 |
|  | 2 | 9 | 4.75 ± 0.92 |
|  | 3 | 8 | 3.12 ± 0.97 |
|  | 4 | 10 | 7.38 ± 1.26 |
| 1 | 1 | 22 | 1.79 ± 2.21 |
|  | 2 | 11 | 6.93 ± 0.41 |
|  | 3 | 21 | 6.75 ± 0.57 |
|  | 4 | 19 | 7.98 ± 0.36 |
| 2 | 1 | 55 | 6.87 ± 1.19 |
|  | 2 | 86 | 1.58 ± 1.24 |
|  | 3 | 79 | 7.98 ± 1.88 |
|  | 4 | 91 | 8.08 ± 0.35 |
| 4 | 1 | 78 | 7.74 ± 0.48 |
|  | 2 | 83 | 8.15 ± 0.28 |
|  | 3 | 79 | 7.59 ± 0.65 |
|  | 4 | 91 | 8.11 ± 0.38 |
| 8 | 1 | 75 | 7.92 ± 0.52 |
|  | 2 | 80 | 7.39 ± 0.67 |
|  | 3 | 80 | 8.04 ± 0.24 |
|  | 4 | 68 | 8.24 ± 0.20 |
| 12 | 1 | 55 | 6.95 ± 0.22 |
|  | 2 | 40 | 6.08 ± 1.57 |
|  | 3 | 34 | 8.27 ± 0.57 |

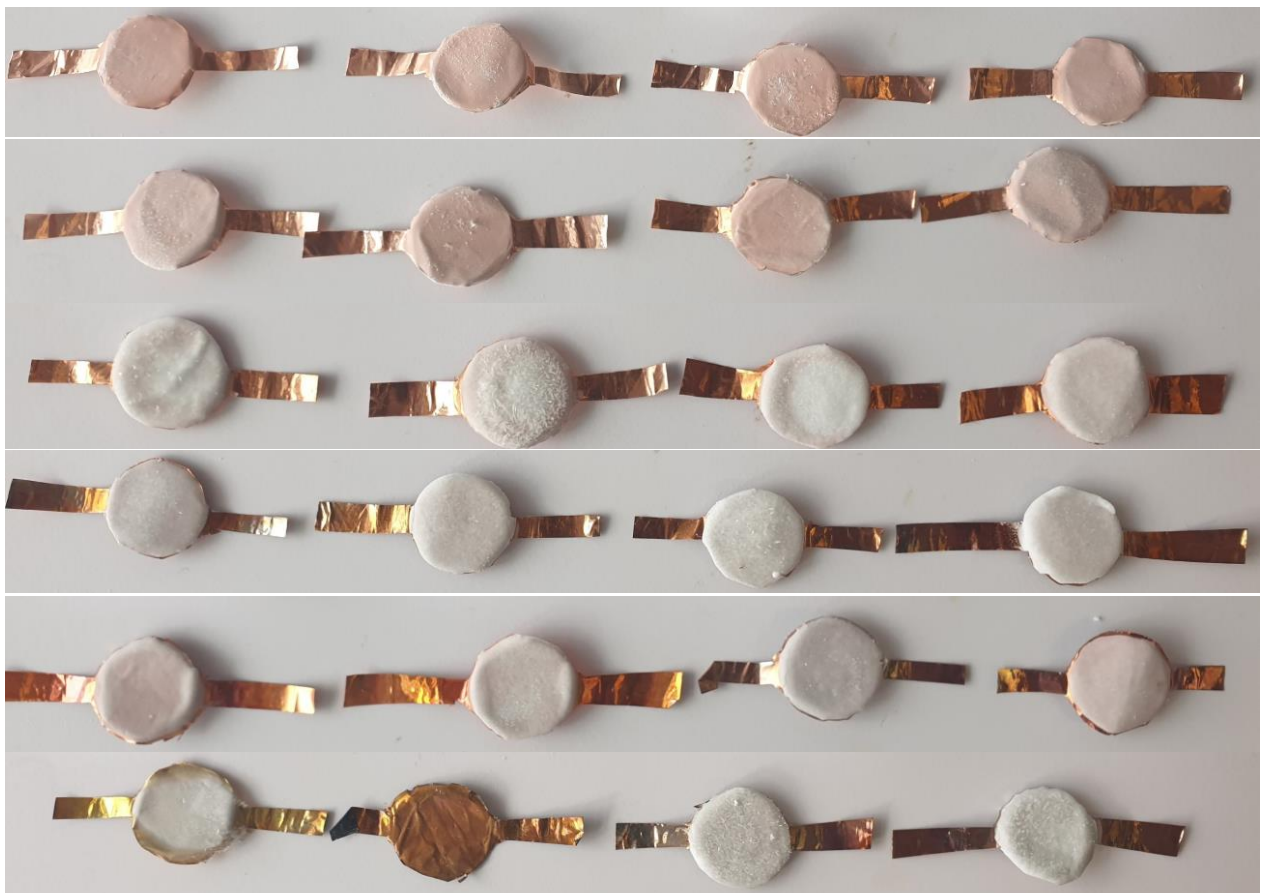

**Figure S13.** Sublimation crystallisation samples (diameter 18mm) prepared at variable time at 200° for 100mg L-Valine. A 30mins, B 1hr, C 2 hrs, D, 4hrs, E 8hrs, F 12hrs

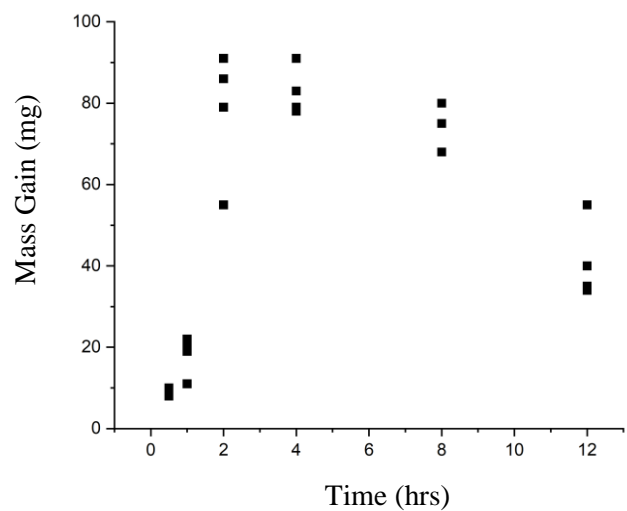

**Figure S14.** Plot of Temperature vs Mass gain for Sublimation samples prepared at variable time at 200°C for 100mg L-Valine.

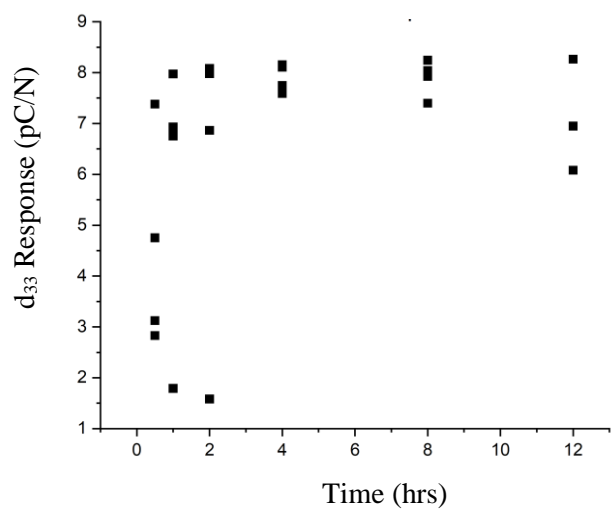

**Figure S15.** Plot of Temperature vs Piezoelectric response for Sublimation samples prepared at variable time at 200°C for 100mg L-Valine.

**Table S6.** Sublimation crystallisation experiments at variable time for at 160°C for 100mg L-Valine

| Time (hrs) | Sample | Weight Gain (mg) | $d_{33}$ Response (pC/N) |
|---|---|---|---|
| 12 | 1 | 10 | 4.45 ± 0.75 |
|  | 2 | 17 | 5.05 ± 0.51 |
|  | 3 | 14 | 6.83 ± 0.35 |
|  | 4 | 20 | 5.87 ± 0.74 |
| 24 | 1 | 40 | 8.02 ± 0.46 |
|  | 2 | 44 | 7.92 ± 0.16 |
|  | 3 | 36 | 7.52 ± 0.27 |
|  | 4 | 37 | 7.78 ± 0.26 |
| 36 | 1 | 59 | 8.59 ± 0.36 |
|  | 2 | 51 | 4.28 ± 1.70 |
|  | 3 | 32 | 7.00 ± 0.69 |
|  | 4 | 27 | 6.63 ± 0.53 |
| 48 | 1 | 77 | 8.87 ± 0.21 |
|  | 2 | 48 | 6.78 ± 1.23 |
|  | 3 | 43 | 7.75 ± 0.27 |
|  | 4 | 69 | 8.23 ± 0.61 |
| 96 | 1 | 52 | 6.89 ± 0.60 |
|  | 2 | 14 | 5.19 ± 0.60 |
|  | 3 | 88 | 7.40 ± 0.85 |
|  | 4 | 74 | 6.57 ± 0.65 |

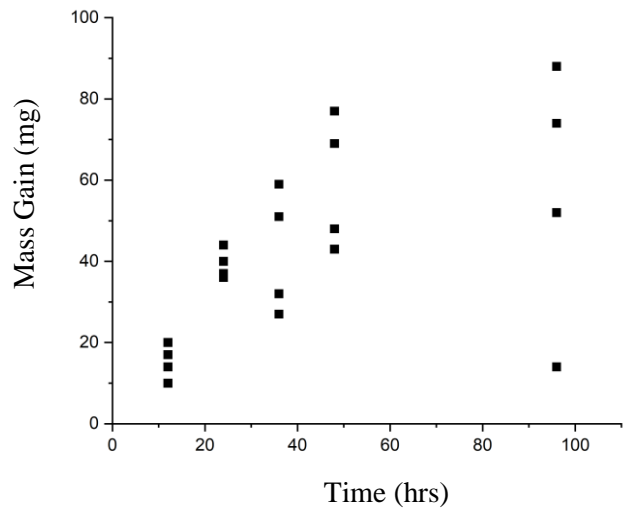

**Figure S16.** Plot of Time vs Mass gain for Sublimation samples prepared at variable time at 160°C for 100mg L-Valine.

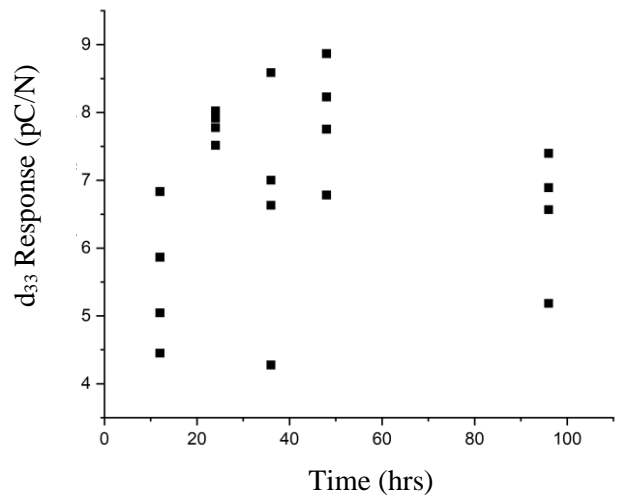

**Figure S17.** Plot of Temperature vs piezoelectric response for Sublimation samples prepared at variable time at 160°C for 100mg L-Valine.

**Table S7.** Sublimation crystallisation experiments at variable temperature for 12 hours for 100mg L-Leucine

| Temperature (°C) | Sample | Weight Gain (mg) | d33 Response (pC/N) |
|---|---|---|---|
| 130 | 1 | 57 | 3.94 ± 0.37 |
|  | 2 | 56 | 3.4 ± 0.6 |
|  | 3 | 53 | 3.35 ± 0.16 |
|  | 4 | 46 | 0.9 ± 0.37 |
| 150 | 1 | 77 | 9.15 ± 0.09 |
|  | 2 | 67 | 9.11 ± 0.09 |
|  | 3 | 69 | - |
|  | 4 | 54 | - |
| 160 | 1 | 40 | 7.16 ± 0.25 |
|  | 2 | 28 | 5.43 ± 0.3 |
|  | 3 | 25 | - |
| 170 | 1 | 78 | - |
|  | 2 | 70 | - |
|  | 3 | 72 | 5.48 ± 0.06 |
|  | 4 | 81 | 3.37 ± 0.21 |
| 180 | 1 | 3 | 0.68 ± 0.18 |

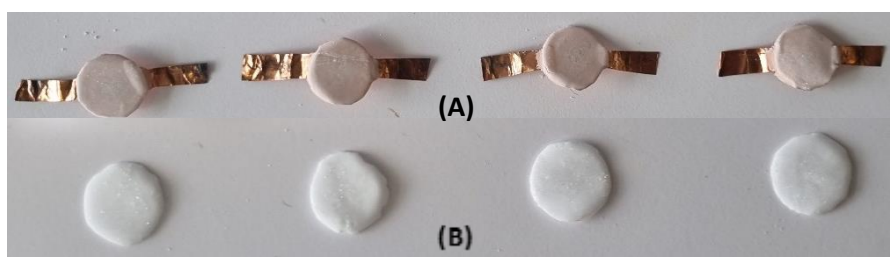

**Figure S18.** Sublimation crystallisation samples (diameter 18mm) prepared for 12hrs at variable time at (a) 130°C, (b) 150°C

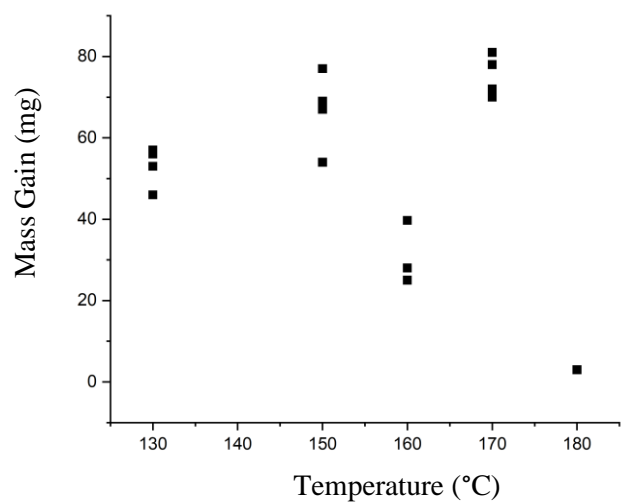

**Figure S19.** Plot of Temperature vs Mass gain for Sublimation samples prepared at variable temperature for 12hrs for 100mg L-Leucine.

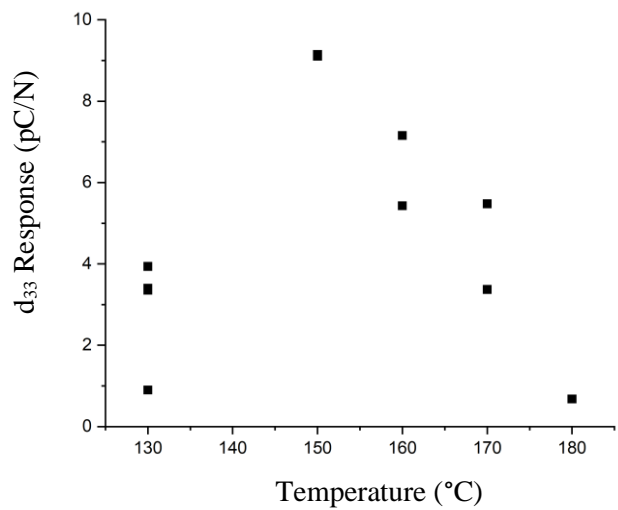

**Figure S20.** Plot of Temperature vs piezoelectric response for Sublimation samples prepared at variable temperature for 12hrs for 100mg L-Leucine.

**Table S8.** Sublimation crystallisation experiments at variable time for at 200°C for 100mg L-Leucine

| Time (hrs) | Sample | Weight Gain (mg) | $d_{33}$ Response (pC/N) |
|---|---|---|---|
| 0.5 | 1 | 82 | 0.44 ± 0.28 |
| | 2 | 69 | 3.93 ± 0.50 |
| | 3 | 77 | - |
| | 4 | 76 | - |
| 1 | 1 | 77 | 3.64 ± 0.27 |
| | 2 | 69 | 4.42 ± 0.47 |
| | 3 | 84 | - |
| | 4 | 60 | - |
| 2 | 1 | 60 | 2.03 ± 0.36 |
| | 2 | 55 | 2.58 ± 0.38 |
| | 3 | 65 | - |
| | 4 | 55 | - |
| 4 | 1 | 69 | 6.92 ± 0.24 |
| | 2 | 72 | 4.21 ± 0.23 |
| | 3 | 79 | 2.90 ± 0.18 |
| | 4 | 71 | - |
| 8+hrs, sample left substrate | | | - |

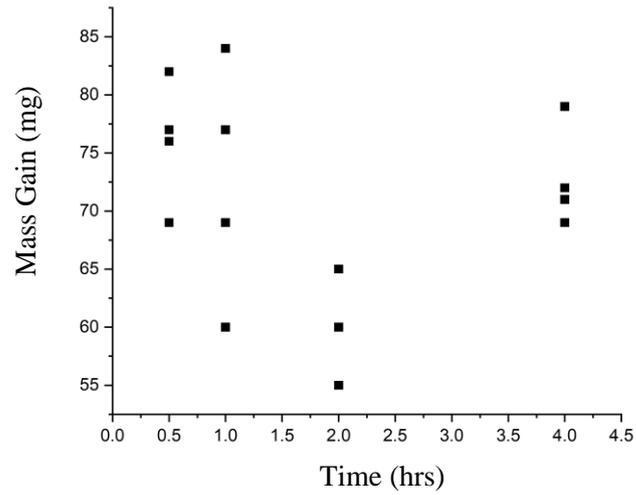

**Figure S21.** Plot of Time vs Mass gain for Sublimation samples prepared at variable time at 200°C for 100mg L-Leucine.

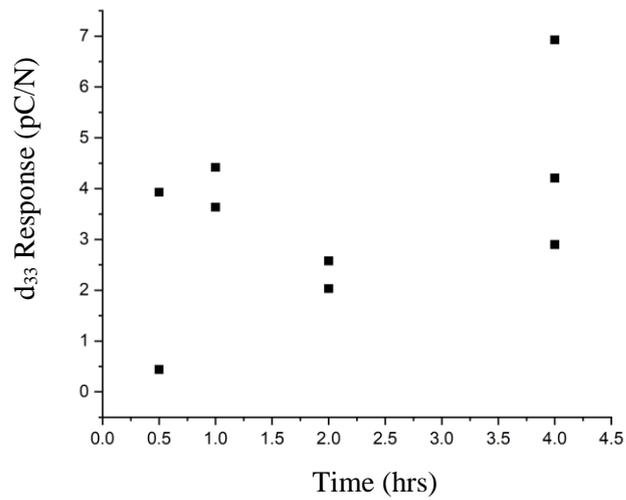

**Figure S22.** Plot of Time vs piezoelectric response for Sublimation samples prepared at variable time at 200°C for 100mg L-Leucine.

**Table S9.** Sublimation crystallisation experiments at variable time for at 160°C for 100mg L-Leucine

| Time (hrs) | Sample | Weight Gain (mg) | $d_{33}$ Response (pC/N) |
|---|---|---|---|
| 2 | 1 | 40 | 7.16 ± 0.25 |
|   | 2 | 28 | 5.43 ± 0.30 |
|   | 3 | 25 | - |
| 4 | 1 | 38 | - |
|   | 2 | 33 | 8.72 ± 0.36 |
|   | 3 | 39 | 3.44 ± 0.21 |
|   | 4 | 27 | - |
| 8 | 1 | 37 | - |
|   | 2 | 56 | - |
|   | 3 | 78 | 5.62 ± 0.83 |
|   | 4 | 68 | 4.51 ± 0.69 |
| 12 | 1 | 83 | 4.88 ± 1.36 |
|   | 2 | 84 | 6.19 ± 0.48 |
|   | 3 | 88 | - |
|   | 4 | 92 | - |
| 24 | 1 | 96 | 6.17 ± 0.71 |
|   | 2 | 95 | 1.10 ± 0.29 |
|   | 3 | 74 | - |
|   | 4 | 64 | - |

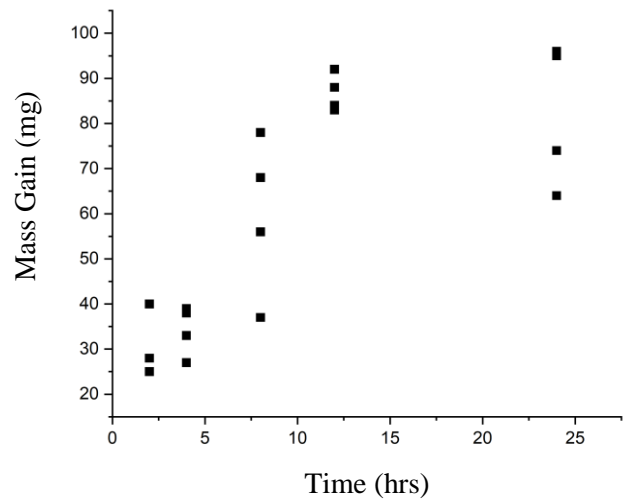

**Figure S23.** Plot of Time vs Mass gain for Sublimation samples prepared at variable time at 160°C for 100mg L-Leucine.

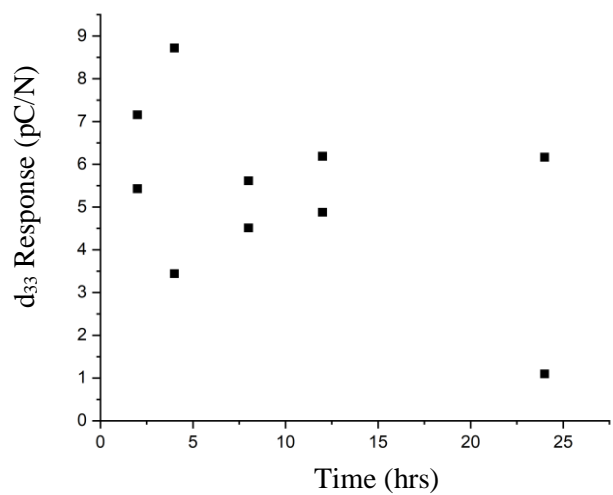

**Figure S24.** Plot of Time vs piezoelectric response for Sublimation samples prepared at variable time at 160°C for 100mg L-Leucine.

**Table S10.** Crystallographic data of amino acid polymorphs

| Crystal | L-Methionine | L-Valine | L-Leucine |
|---|---|---|---|
| CCDC Ref. code | LMETOB12 | LVALIN05 | LEUCIN02 |
| Formula | $C_5H_{11}NO_2S$ | $C_5H_{11}NO_2$ | $C_6H_{13}NO_2$ |
| $M_r$ | 187.794 | 153.808 | 183.491 |
| Crystal System | Monoclinic | Monoclinic | Monoclinic |
| Space Group | *P*21 | *P*21 | *P*21 |
| Unit Cell Dimensions | | | |
| a (Å) | 9.512(1) | 9.670(1) | 9.562(2) |
| b (Å) | 5.194(1) | 5.275(1) | 5.301(1) |
| c (Å) | 15.342(1) | 12.063(2) | 14.519(3) |
| α (°) | 90.00 | 90.00 | 90.00 |
| β (°) | 97.64(1) | 90.80(1) | 94.20(2) |
| γ (°) | 90.00 | 90.00 | 90.00 |
| V (Å$^3$) | 751.178 | 615.233 | 183.491 |
| $D_{calc}$ (g/cm$^3$) | 1.319 | 1.265 | 1.187 |
| Z | 4 | 4 | 4 |

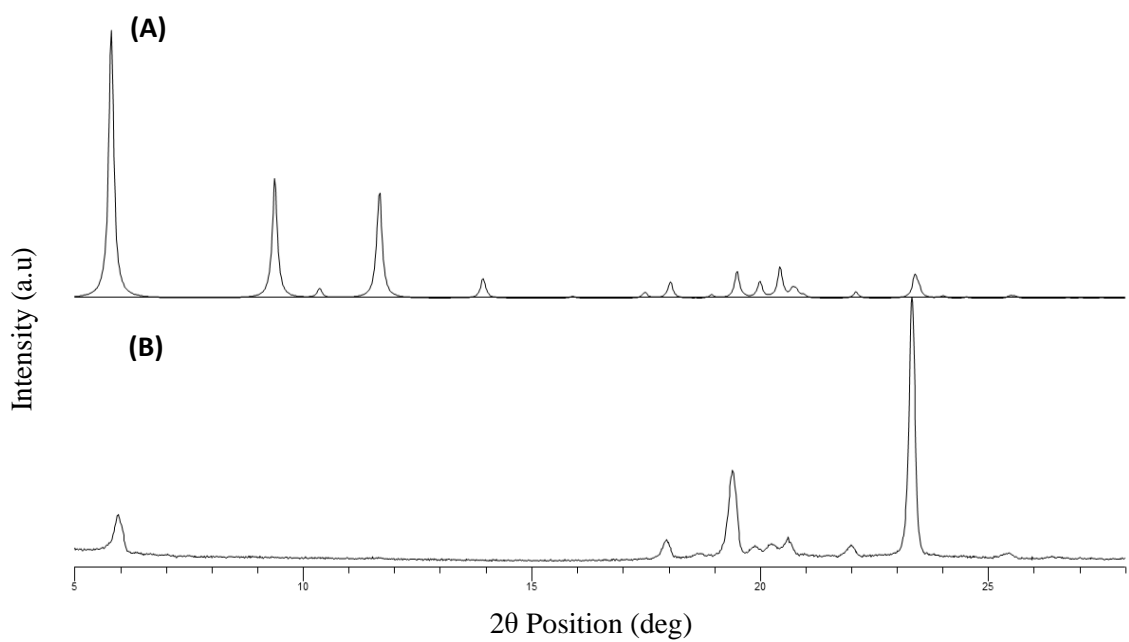

**Figure S25.** Theoretical XRPD pattern of L-Methionine calculated from the single crystal data (A) compared with sublimated L-Methionine (B)

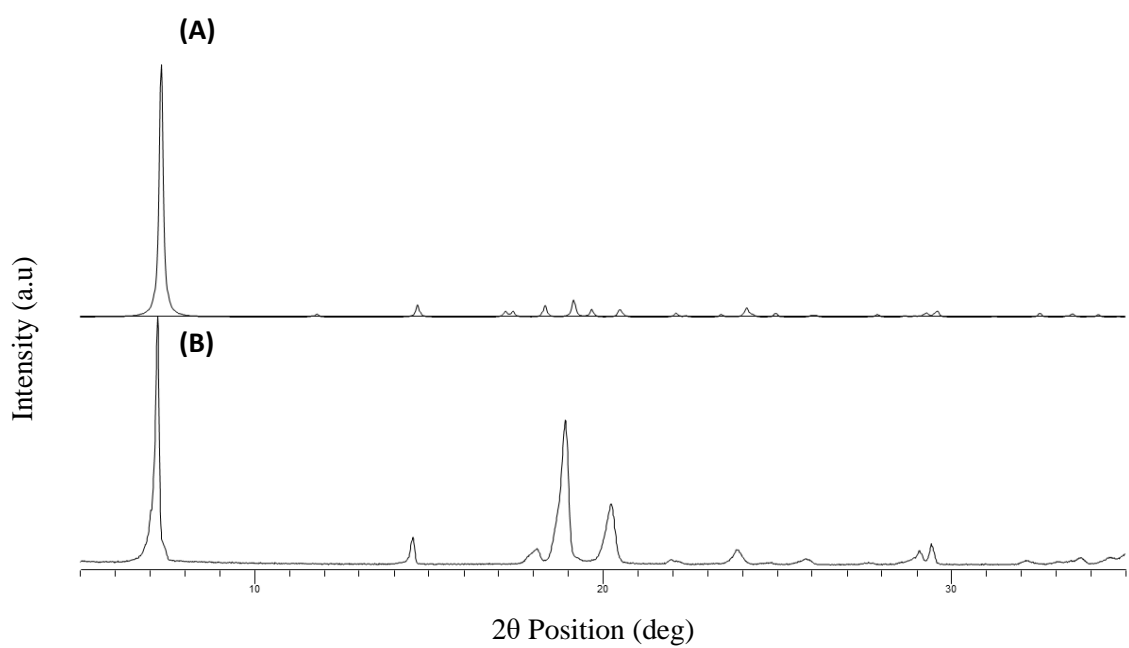

**Figure S26.** Theoretical XRPD pattern of L-Valine calculated from the single crystal data (A) compared with sublimated L-Valine (B)

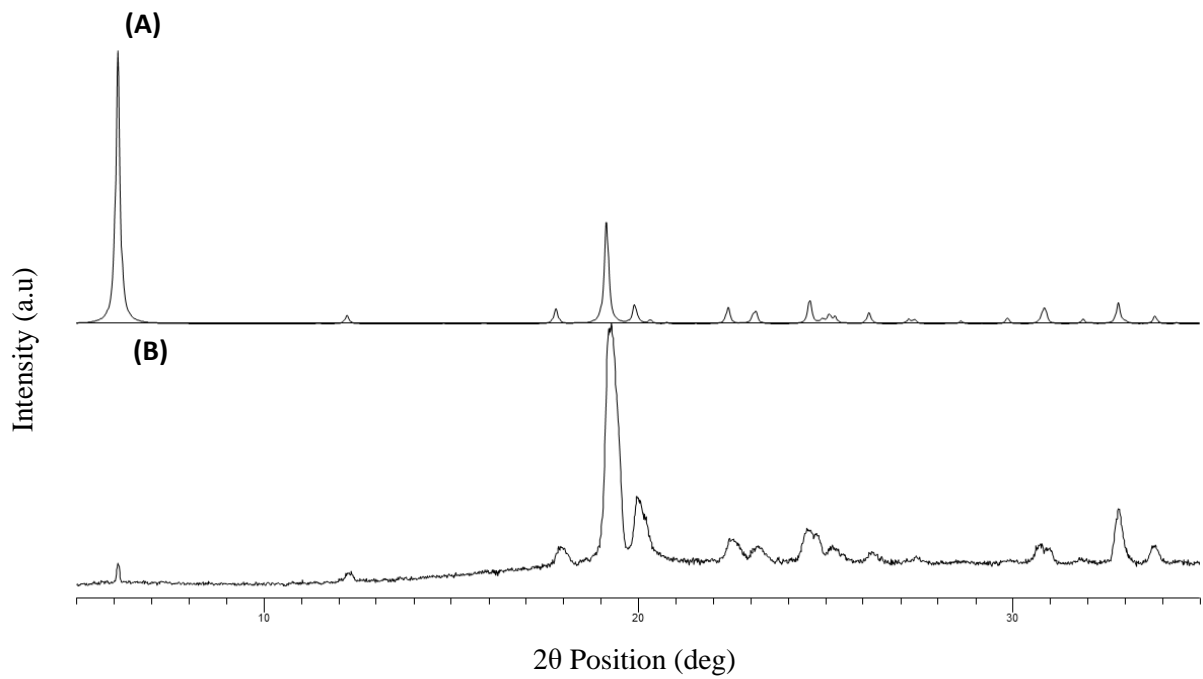

**Figure S27**. Theoretical XRPD pattern of L-Leucine calculated from the single crystal data (A) compared with sublimated L-Leucine (B)

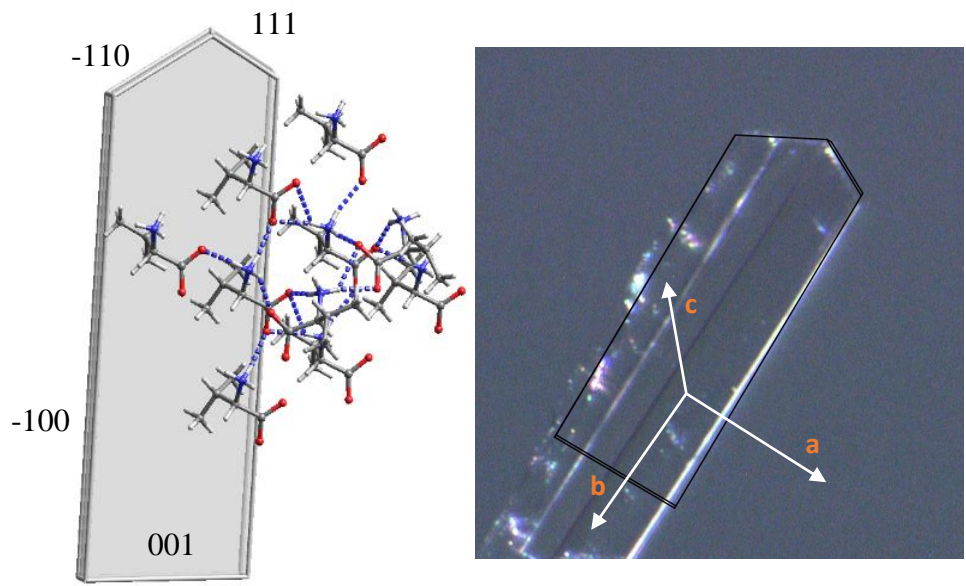

**Figure S28.** Experimental morphology characterisation of L-Valine from sublimation.

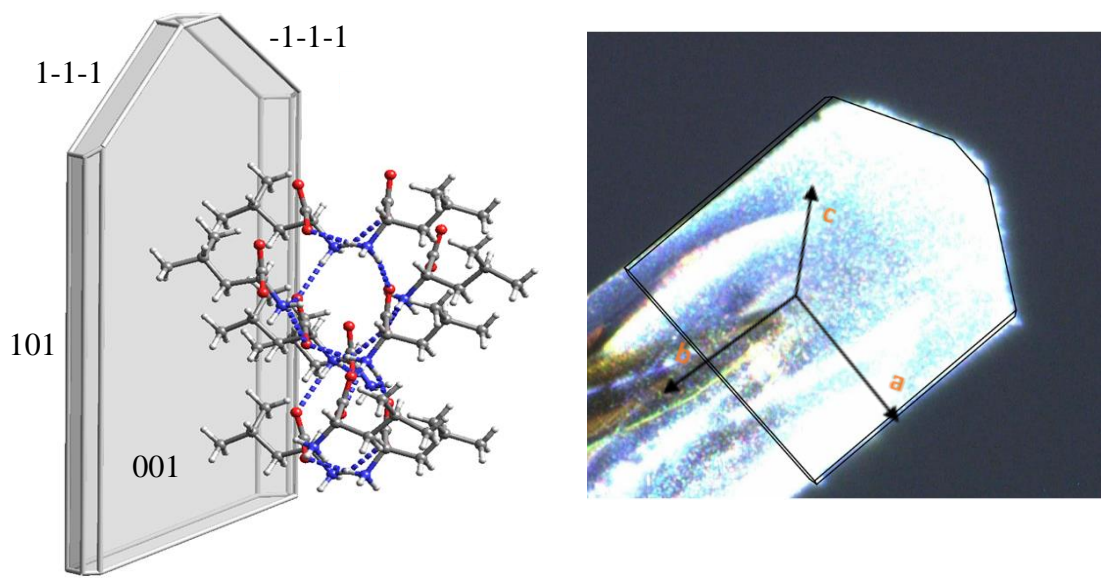

**Figure S29.** Experimental morphology characterisation of L-Leucine from sublimation

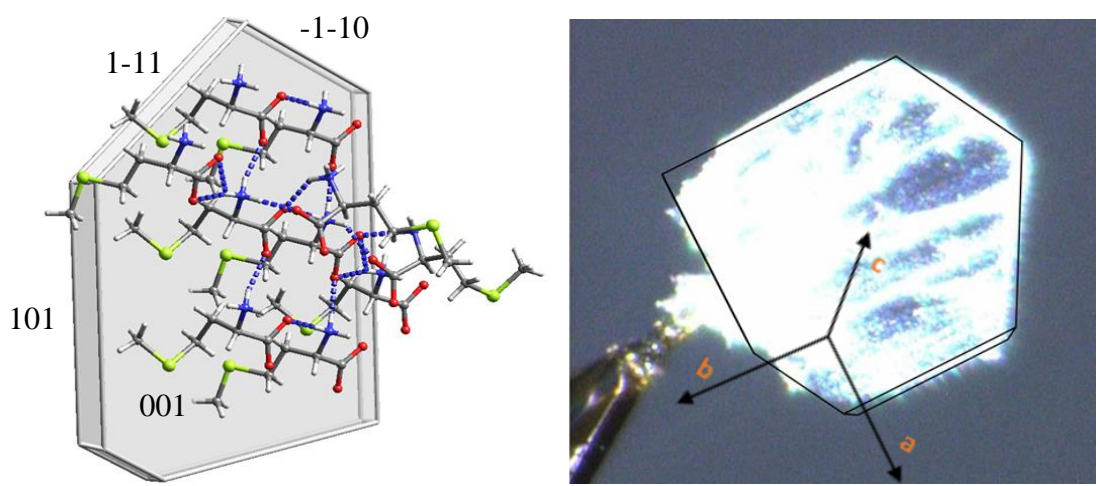

**Figure S30**. Experimental morphology characterisation of L-Methionine from sublimation

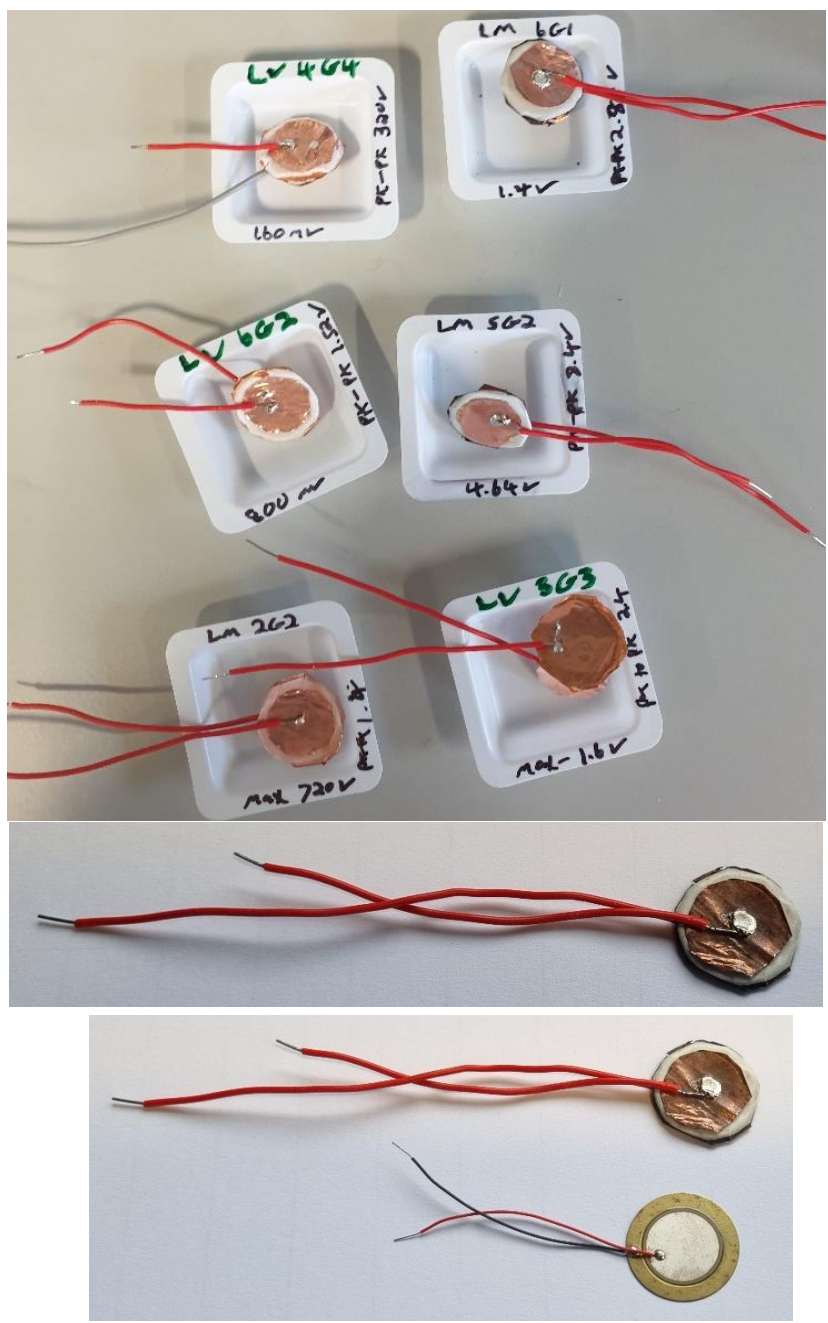

**Figure S31. EXAMPLES** Samples prepared for Voltage testing of L-Valine and L-Methionine. / compared to a commercial piezoelectric buzzer

**Table S11.** Voltage testing of samples of L-Valine and L-Methionine

| Sample | Temperature (°C) | Time (hrs) | Weight Gain (mg) | Average d33 | Std Dev | Max V | Pk-Pk V |
|---|---|---|---|---|---|---|---|
| L-Valine 1 | 160 | 36 | 32 | 7.002 | 0.696 | 1.6 | 2.4 |
| L-Valine 2 | 160 | 48 | 69 | 8.23 | 0.612 | 0.16 | 0.32 |
| L-Valine 3 | 160 | 96 | 14 | 5.185 | 0.598 | 0.8 | 1.52 |
| L-Methionine 4 | 200 | 30 | 6 | 4.112 | 0.357 | 0.72 | 1.4 |
| L-Methionine 5 | 200 | 4 | 47 | 0.892 | 0.314 | 4.64 | 8.4 |
| L-Methionine 6 | 200 | 8 | 66 | 0.617 | 0.267 | 1.4 | 2.8 |

**Table S12.** Voltage testing of samples of L-Methionine prepared at 160°C for 12 hours

| Sample | Mass Gain (mg) | d33 response (pC/N) | Max V (V) | Pk-Pk V (V) |
|---|---|---|---|---|
| 1 | 4 | 6.3 | 0.64 | 1.54 |
| 2 | 4 | 5.8 | 0.8 | 1.60 |
| 3 | 6 | 3.2 | - | - |
| 4 | 6 | 5.5 | 0.64 | 1.12 |
| 5 | 4 | 5.3 | - | - |
| 6 | 4 | 4.6 | 0.62 | 1.32 |
| 7 | 5 | 5.6 | 0.56 | 1.92 |
| 8 | 4 | 4.7 | 0.72 | 1.52 |

**Table S13.** Voltage testing of samples of L-Methionine prepared at 200°C for 4 hours

| Sample | Mass Gain (mg) | d33 response (pC/N) | Max V (V) |
|---|---|---|---|
| 1 | 90 | 5.82 | - |
| 2 | 84 | 5.57 | 1.277 |
| 3 | 91 | 5.80 | 1.612 |
| 4 | 86 | 5.68 | 1.428 |
| 5 | 87 | 5.99 | 1.419 |
| 6 | 89 | 5.98 | 1.577 |
| 7 | 89 | 5.39 | 1.293 |
| 8 | 87 | 5.42 | 1.599 |
|  | Average | 5.71 |  |

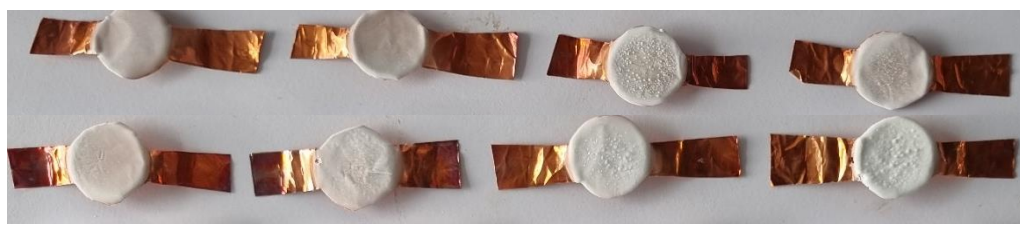

**Figure S32.** Samples prepared for Voltage testing L-Methionine from sublimation for 4hrs at 200°C.

**Table S14.** Thickness Measurements of L-Leucine samples prepared by sublimation

| Temperature (°C) | Time (hrs) | Sample | Weight Gain (mg) | d33 Response (pC/N) | Average Thickness (mm) | Average Thickness – Average Thickness of substrate (mm) |
|---|---|---|---|---|---|---|
| 130 | 12 | 1 | 57 | 3.94 ± 0.37 | 0.639 ± 0.079 | 0.515 ± 0.083 |
| 130 | 12 | 3 | 53 | 3.35 ± 0.16 | 0.479 ± 0.051 | 0.355 ± 0.055 |
| 130 | 12 | 4 | 46 | 0.9 ± 0.37 | 0.452 ± 0.058 | 0.328 ± 0.062 |
| 150 | 12 | 1 | 77 | 9.15 ± 0.09 | 0.330 ± 0.079 | 0.206 ± 0.083 |
| 150 | 12 | 2 | 67 | 9.11 ± 0.09 | 0.547 ± 0.127 | 0.423 ± 0.131 |
| 160 | 2 | 1 | 40 | 7.16 ± 0.25 | 0.131 ± 0.039 | 0.007 ± 0.043 |
| 160 | 4 | 1 | 38 | - | 0.397 ± 0.023 | 0.273 ± 0.027 |
| 160 | 8 | 4 | 68 | 4.51 ± 0.69 | 0.780 ± 0.036 | 0.656 ± 0.04 |
| 160 | 12 | 4 | 92 | - | 0.983 ± 0.023 | 0.859 ± 0.027 |
| 160 | 24 | 1 | 96 | 6.17 ± 0.71 | 0.831 ± 0.062 | 0.707 ± 0.066 |
| 170 | 12 | 1 | 78 | - | 0.524 ± 0.031 | 0.4 ± 0.035 |
| 170 | 12 | 4 | 81 | 3.37 ± 0.21 | 0.627 ± 0.052 | 0.503 ± 0.056 |

**Table S15.** Thickness Measurements of L-Methionine samples prepared by sublimation

| Temperature (°C) | Time (hrs) | Sample | Weight Gain (mg) | d33 Response (pC/N) | Average Thickness (mm) | Average Thickness – Average Thickness of substrate (mm) |
|---|---|---|---|---|---|---|
| 160 | 12 | 2 | 6 | 6.57 ± 0.66 | 0.217 ± 0.074 | 0.093 ± 0.078 |
| 170 | 12 | 2 | 10 | 4.79 ± 0.14 | 0.138 ± 0.004 | 0.014 ± 0.008 |
| 180 | 12 | 1 | 16 | 5.13 ± 0.10 | 0.162 ± 0.008 | 0.038 ± 0.012 |
| 190 | 12 | 1 | 47 | 4.51 ± 0.31 | 0.224 ± 0.036 | 0.1 ± 0.04 |
| 200 | 12 | 2 | 68 | 3.31 ± 0.61 | 0.313 ± 0.018 | 0.189 ± 0.022 |
| 200 | 0.5 | 1 | 6 | 4.11 ± 0.36 | 0.143 ± 0.008 | 0.019 ± 0.012 |
| 200 | 1 | 4 | 10 | 4.25 ± 0.29 | 0.149 ± 0.01 | 0.025 ± 0.014 |
| 200 | 2 | 3 | 22 | 3.61 ± 0.41 | 0.434 ± 0.019 | 0.31 ± 0.023 |
| 200 | 4 | 1 | 48 | 5.78 ± 0.35 | 0.906 ± 0.055 | 0.782 ± 0.059 |
| 200 | 8 | 3 | 36 | 4.76 ± 0.23 | 0.68 ± 0.166 | 0.556 ± 0.17 |